\documentclass[nofootinbib,aps,prd,groupedaddress,preprintnumbers,%
  showpacs,showkeys,floatfix,amssymb,amsfonts]{revtex4-1}  
\usepackage{hyperref}
\usepackage[usenames]{color}
\usepackage{todonotes}
\usepackage{bbold}
\usepackage{amsmath}
\usepackage{multirow}
\usepackage{graphicx}
\usepackage{xfrac}
\usepackage[utf8]{inputenc}

\usepackage{amsfonts}
\usepackage{amssymb}
\usepackage{extarrows}

\newcommand{\be}{\begin{equation}}
\newcommand{\ee}{\end{equation}}
\newcommand{\bea}{\begin{eqnarray}}
\newcommand{\eea}{\end{eqnarray}}

\newcommand{\Op}{{\cal O}}
\newcommand{\avgx}{\langle x \rangle}
\newcommand{\avgxx}{\langle x^2 \rangle}
\newcommand{\Dlr}{\buildrel \leftrightarrow \over D\raise-1pt\hbox{}}

\newcommand{\eins}{\hat{1}} 

\newcommand{\csw}{c_\mathrm{sw}}

\newcommand{\avgxxx}{\langle x^3 \rangle}
\newcommand{\avgxxxx}{\langle x^4 \rangle}
\newcommand{\avgxxxxx}{\langle x^5 \rangle}
\newcommand{\avgxxxxxx}{\langle x^6 \rangle}

\newcommand{\J}{\mathcal{J}} 
\newcommand{\s}{{\mathcal S}} 

\newcommand{\emdash}{\hspace{1pt}---\hspace{1pt}}

\begin{document}

\title{The pion and kaon $\langle x^3 \rangle$ from lattice QCD and \\ PDF reconstruction from Mellin moments}

\author{
\vspace*{0.35cm}
  Constantia Alexandrou$^{1,2}$,
  Simone Bacchio$^2$,
  Ian Clo\"et$^3$,
  Martha Constantinou$^{4}$,\\[1ex]
Kyriakos Hadjiyiannakou$^{1,2}$,
  Giannis Koutsou$^{2}$,
  Colin Lauer$^{3,4}$\\[2ex]
 (ETM Collaboration)
}

\affiliation{
  \vskip 0.25cm
  $^1$Department of Physics, University of Cyprus,  P.O. Box 20537,  1678 Nicosia, Cyprus\\
  \vskip 0.05cm
    $^2$Computation-based Science and Technology Research Center,
  The Cyprus Institute, 20 Kavafi Str., Nicosia 2121, Cyprus \\
  \vskip 0.05cm
  $^3$ Physics Division, Argonne National Laboratory, Lemont, Illinois 60439, USA\\
  \vskip 0.05cm
  $^4$Department of Physics, Temple University, 1925 N. 12th Street, Philadelphia, PA 19122-1801, USA\\ 
  \vskip 0.05cm
\phantom{-}
\centerline{\today}
 {\includegraphics[scale=0.2]{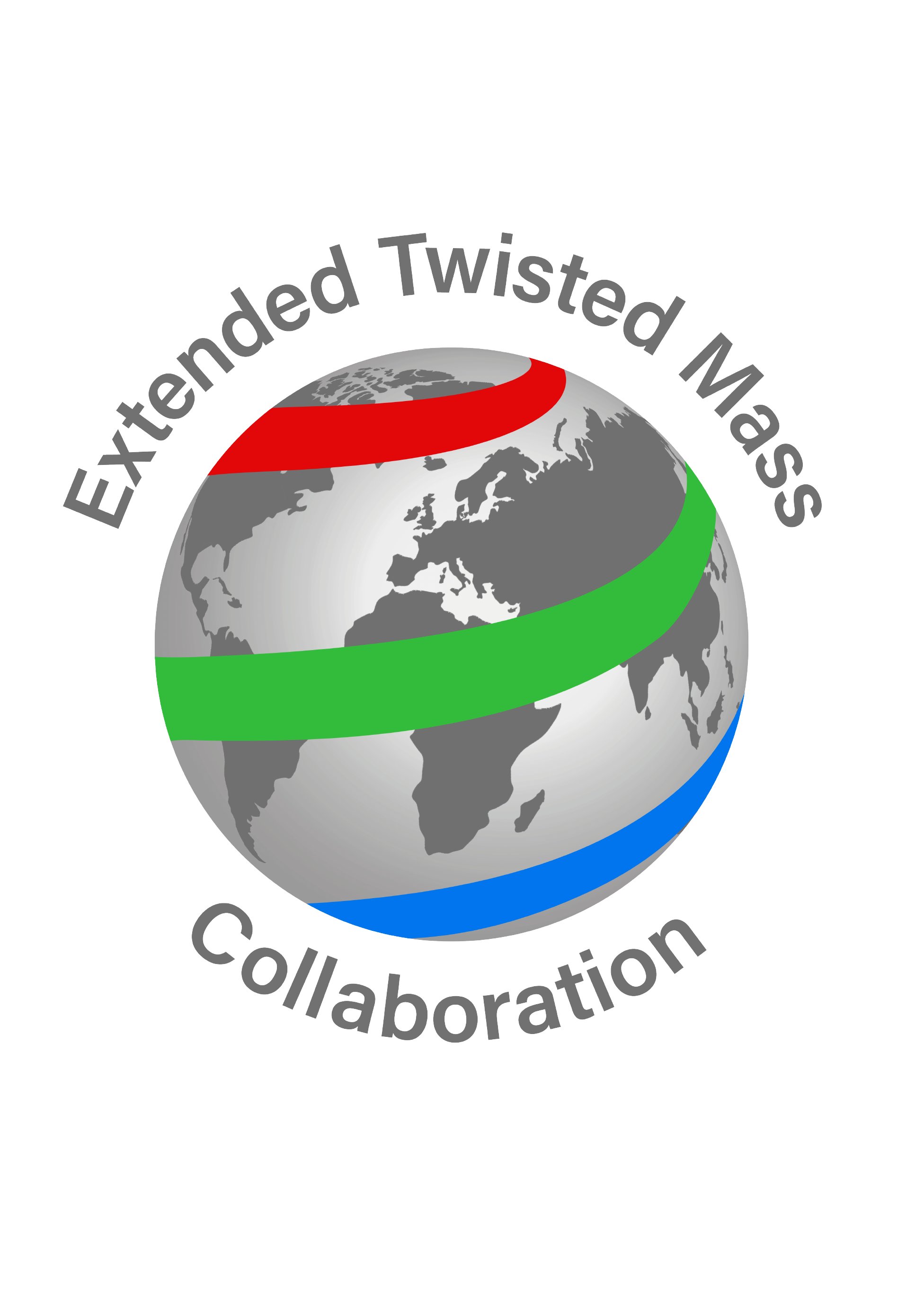}}
  }


\begin{abstract}
  We present a calculation of the pion and kaon Mellin moment $\langle x^3 \rangle$ extracted directly in lattice QCD using a three-derivative local operator. We use one ensemble of gauge configurations with two degenerate light, a strange and a charm quark ($N_f=2+1+1$) of maximally twisted mass fermions with clover improvement. The ensemble reproduces a pion mass $\sim260$ MeV, and a kaon mass $\sim530$ MeV. Excited-states contamination is evaluated using four values of the source-sink time separation within the range of $1.12-1.67$~fm. We use an operator that is free of mixing, and apply a multiplicative renormalization function calculated non-perturbatively. Our results are converted to the $\overline{\rm MS}$ scheme and evolved at a scale of 2 GeV, using three-loop expressions in perturbation theory. The final values are $\langle x^3 \rangle_\pi^{u^+}=0.024(18)_{\rm stat}(2)_{\rm syst}$, $\langle x^3 \rangle_K^{u^+}=0.035(6)_{\rm stat}(3)_{\rm syst}$, and $\langle x^3 \rangle_K^{s^+}=0.075(5)_{\rm stat}(1)_{\rm syst}$, where the systematic error is the uncertainty due to excited state contamination.

We combine $\langle x^3 \rangle$ with the two lower moments, namely $\langle x \rangle$ and $\langle x^2 \rangle$, to obtain the ratios $\langle x^3 \rangle/\langle x \rangle$ and $\langle x^3 \rangle/\langle x^2 \rangle$, as well as $\langle x^3 \rangle^{u^+}_K/ \langle x^3 \rangle^{u^+}_\pi$ and $\langle x^3 \rangle^{u^+}_K/ \langle x^3 \rangle^{u^+}_\pi$.  In addition, we reconstruct the $x$-dependence of the pion and kaon PDFs via 2- and 3-parameter fits to our results. We find that the reconstruction is feasible and  that our lattice data favor a large $x$-dependence that falls as $(1-x)^2$ for both the pion and kaon PDFs. We integrate the reconstructed PDFs to extract the higher moments with $4\leq n\leq 6$. Finally, we compare the pion and kaon PDFs, as well as the ratios of their moments, to address the effect of SU(3) flavor symmetry breaking.

\end{abstract}

\maketitle

\section{Introduction}
The pion, kaon, and eta mesons comprise the octet of Nambu--Goldstone bosons, which are unique among hadrons because their masses vanish in the chiral limit. The valence quark structure of these mesons is given by a combination of a quark and an anti-quark with flavors $u$, $d$, and $s$, and if the masses of these quarks are equal then these Nambu--Goldstone bosons are mass degenerate (up to electroweak effects). Since the mass of the strange quark is significantly higher than that of the light $u$ and $d$ quarks\emdash $2\,m_s/(m_u + m_d) = 27.46 \pm 0.15 \pm 0.41$
~\cite{Zyla:2020zbs}\emdash comparison between pion and kaon observables provides a unique window into the interplay between strong interaction forces described by quantum chromodynamics (QCD) and quark mass effects~\cite{Hutauruk:2016sug}.

For the pion and kaon, signficant SU(3) flavor breaking effects have already been observed. For example, experiment finds pion and kaon charge radii of $r_{\pi^+} = 0.672 \pm 0.008$~fm and  $r_{K^+} = 0.560 \pm 0.031$~fm~\cite{Zyla:2020zbs}, which reveals flavor breaking effects of around 10\%. More striking perhaps is for the neutral pion and kaon, where $r_{\pi^0}$ vanishes and $r_{K^0} = -0.277 \pm 0.018$~fm~\cite{Zyla:2020zbs}. Similar effects are expected in the pion and kaon parton distribution functions (PDFs). However, while some data exists for the pion, from pion induced Drell-Yan~\cite{Conway:1989fs}, knowledge of the kaon is even more limited with only some early data on $\bar{u}_{K^-}(x)/\bar{u}_{\pi^-}(x)$~\cite{Badier:1983mj}. The naive expectation based on quark mass effects is that $s_K(x)$ will have more support at large $x$ (harder) while $u_K(x)$ will be concentrated at smaller $x$ (softer). Similarly, flavor breaking would imply $s_K(x)$ is harder and $u_K(x)$ is softer than the $u/d$ quark distributions in the pion. Existing data on the pion and kaon are not in contradiction to these naive expectations, however, the error bars are large so definitive conclusions cannot be made.

Calculations of the pion and kaon are also limited within lattice QCD as compared to the proton. Among the first calculations for the pion are for the moments in the quenched approximation~\cite{Best:1997qp}, which were later improved~\cite{Brommel:2006zz,Brommel:2007zz}. It is only recently that the first calculation of moments using local operators was performed~\cite{Alexandrou:2020gxs}. Lattice calculations of the $x$-dependence of the pion and kaon PDFs became available in the last few years~\cite{Karpie:2018zaz,Sufian:2019bol,Izubuchi:2019lyk,Joo:2019bzr,Bali:2019dqc,Lin:2020ssv,Gao:2020ito,Sufian:2020vzb} using methods like the quasi-PDFs~\cite{Ji:2013dva,Ji:2014gla}, pseudo-Ioffe-time-distributions (ITD)~\cite{Radyushkin:2016hsy}, and current-current correlators~\cite{Ma:2014jla,Ma:2014jga,Ma:2017pxb}. For a recent review on these approaches see Refs.~\cite{Cichy:2018mum,Constantinou:2020pek}. Using these methods, one can integrate and extract the $n^{\rm th}$ moments.

In this work, we calculate the non-trival moments of the pion and kaon quark PDFs up to $\left<x^3\right>$ using lattice QCD, and explore the size of quark mass effects by comparing moments in the pion and kaon, and between the light and strange quarks in the kaon. These moments are determined by directly evaluating the associated operators, using one ensemble of gauge configurations with two degenerate light quarks, and strange and charm quarks ($N_f = 2 + 1 + 1$) of maximally twisted mass fermions with a clover improvement. We avoid operator mixing in the $\left<x^3\right>$ moments by using three different spatial and a temporal component for the associated operator. The computation of these moments provide important insight into the large-$x$ behavior of the pion and kaon PDFs, and provide a check of systematic errors associated with recent methods to determine the full $x$-dependence of the PDFs using, e.g., quasi-PDFs~\cite{Ji:2013dva,Gao:2020ito}, pseudo-PDFs~\cite{Radyushkin:2017cyf,Joo:2020spy}, and current-current correlation methods~\cite{Ma:2017pxb,Sufian:2020vzb}. With $\left<x^3\right>$ at hand for the pion and kaon, and the lower two moments calculated in Ref.~\cite{Alexandrou:2020gxs}, we make an attempt to reconstruct the $x$-dependence of the PDFs by fitting to a functional form, which is constrained by the baryon number sum rule. Sensitivity to the exponent $\beta$ of the reconstructed PDFs $q(x) \simeq (1-x)^\beta$ near $x \sim 1$ is explored. Furthermore, we discuss the size of SU(3) flavor breaking effects by comparing the pion and kaon moments as well as the reconstructed PDFs.

The paper is organized as follows: In Sec.~\ref{sec:II}, we present the theoretical framework and the decomposition to obtain the $\left<x^3\right>$ quark moments for the pion and kaon. We refer the reader to Ref.~\cite{Alexandrou:2020gxs} for the formalism associated with the $\left<x\right>$ and $\left<x^2\right>$ quark moments. In Sec.~\ref{sec:III} we describe how to perform the non-perturbative renormalization of the three-derivative operator associated with $\left<x^3\right>$ and in Sec.~\ref{sec:analysis} we provide details on the analysis methods used to obtain the pion and kaon moments. Our results for the pion and kaon quark PDF moments are presented in Sec.~\ref{sec:fin_results}, where comparisons are made to other lattice QCD results and various model calculations. In Sec.~\ref{sec:PDFs}, we present the reconstruction of the $x$-dependence of the PDFs and in Sec.~\ref{sec:VII} we summarize this work.

\section{Theoretical and Lattice Setup\label{sec:II}}

The meson matrix elements connected to $\avgxxx$, $\langle M | \Op^{\{\mu\nu\rho\sigma \}} | M \rangle$, with $|M\rangle$ being a meson state, contain a bilinear fermion vector operator with three covariant derivatives, that is
\begin{equation}
    \Op^{\{\mu\nu\rho\sigma\}} \equiv \overline{\psi} \gamma^{\{\mu} D^\nu D^\rho D^{\sigma\}} \psi \,,
\end{equation}
where $\psi$ is a quark field. Curly brackets denote symmetrization over the indices and subtraction of the trace. In general, any of the indices $\mu,\,\nu,\,\rho,\,\sigma$ can be spatial (1,2,3) or temporal (4) and in any combination. However, to avoid mixing with other operators we choose all four indices to be different from each other~\cite{Capitani:1994qn,Beccarini:1995iv,Gockeler:1996mu,Gockeler:2004wp}, which leads to the operator $\Op^{\{1234\}}$. There are twenty-four permutations for the indices, which are all calculated and averaged over.

The relevant decomposition of the meson matrix element of the operator $\Op^{\{\mu\nu\rho\sigma \}}$ in Euclidean space leads to three generalized form factors, via the expression
\begin{eqnarray}
\label{decomp}
\hspace*{-0.25cm}
\langle M (p') | \Op^{\{\mu\nu\rho\sigma \}} | M (p) \rangle = 
C\left[
-2 P^{\{\mu} P^{\nu} P^{\rho}  P^{\sigma\}} \,A_{40}(Q^2) 
-2\Delta^{\{\mu} \Delta^{\nu} P^{\rho}P^{\sigma\}}\,A_{42}(Q^2) 
-2\Delta^{\{\mu} \Delta^{\nu} \Delta^{\rho} \Delta^{\sigma\}}\,A_{44}(Q^2) \right],\,\,
\end{eqnarray}
in a general frame with initial momentum $p$ and final momentum $p'$. $P$ is defined as the average of the initial and final momenta of the meson, $P=(p+p')/2$, and $\Delta$ is their difference, $\Delta=p'-p$. The generalized form factors $A_{ij}$ are only dependent on the momentum transferred squared, $Q^2$. $C$ is a kinematic factor, which depends on the normalization of the meson state. In this work, we obtain $C=\frac{1}{\sqrt{4 E(p) E(p')}}$ for a general frame, where $m_M$ is the mass of meson $M$ and $E(p){=}\sqrt{m_M^2 + \vec{p}\,^2}$ is the energy at momentum $\vec{p}$. 

To extract the moment $\avgxxx\equiv A_{40}(0)$, we study the forward-kinematics limit of Eq.~(\ref{decomp}), that is, $p'=p \ne 0$,\,$Q^2=0$, which leads to
\begin{eqnarray}
\langle M(p)| \Op^{\{\mu\nu\rho\sigma \}} |M(p)\rangle &=& -2 \frac{1}{2E}\, p^{\{\mu} \,p^\nu\, p^\rho\, p^{\sigma\}}\, \avgxxx_M\,.
\end{eqnarray}
The kinematic factor $C$ is simplified to $1/(2E)$. Similarly to $\avgxx$, the kinematic coefficient of $\avgxxx$ becomes zero in the rest frame ($\vec{p}\,'=\vec{p}=0$) at the forward limit, unless all the indices of the operator are temporal. However, the renormalization of ${\cal O}^{4444}$ is very complicated as it is not multiplicative and requires one to disentangle $\avgxxx$ from with lower-dimension operators~\cite{Capitani:1994qn,Beccarini:1995iv,Gockeler:1996mu}. The only possibility to avoid any 
mixing in $\avgxxx$ is to use the operator ${\cal O}^{\{1234\}}$, which we use in this work, and obtain:
\begin{eqnarray}
\label{decomp2}
\langle M(p)| \Op^{\{1234\}} |M(p)\rangle &=& -i\, p^1\, p^2\, p^3\, \avgxxx_M\,.
\end{eqnarray}
We employ the simplest boosted frame setup to get non-vanishing matrix elements, that is, all spatial components of the four-momentum are equal, $p'=p=(i E,p^1,p^2,p^3)$. 
To have the highest possible signal-to-noise ratio in this setup we choose the smallest nonzero value for the spatial components, that is $\vec{p}\,'=\vec{p} = \frac{2\pi}{L}(\pm1,\pm1,\pm1)$, where $L$ is the spatial extend of the lattice used. To increase statistics, we use all eight permutations of the momentum boost, and all twenty-four combinations of indices that enter the symmetrization of ${\cal O}^{\{1234\}}$.

\begin{figure}[h]
    \centering
    \includegraphics[scale=0.25]{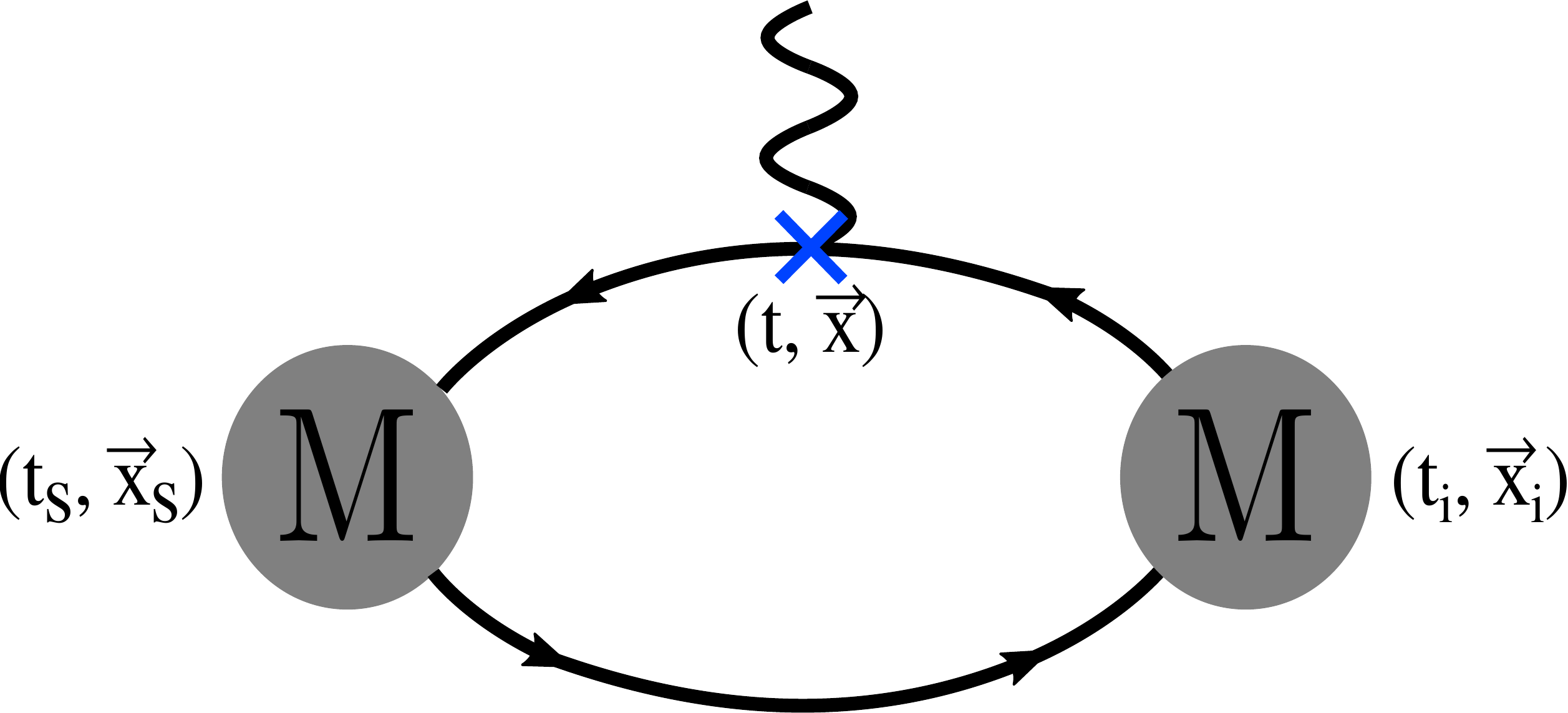}
\caption{Connected diagram for the three-point function entering the calculation of $\langle x \rangle$ and $\avgxx$. The wavy line corresponds to the operator insertion.}
\label{fig:3ptDiagram}
\end{figure}

\medskip
We calculate the connected contributions to $\avgxxx$, shown in the pictorial representation of Fig.~\ref{fig:3ptDiagram}. We use an ensemble of gauge configurations labeled cA211.30.32, which has been produced by the Extended Twisted Mass Collaboration (ETMC)~\cite{Alexandrou:2018egz}. The ensemble uses $N_f=2+1+1$ twisted clover fermions with clover improvement and the Iwasaki improved gluon action. The fermionic action is written as the sum of the degenerate light ($S_{tm}^\ell$) and non-degenerate heavy ($S_{tm}^h$) fermion actions  
\begin{equation}
  S_{tm}^\textrm{ferm} = S_{tm}^\ell + S_{tm}^h,
  \label{eq:Sferm}
\end{equation}
with:
\begin{align}
  \label{eq:sflh}
  S_{tm}^\ell =& \sum_x \bar{\chi}_\ell(x)\left[ D_W(U) + \frac{i}{4} \csw \sigma^{\mu\nu}
    \mathcal{F}^{\mu\nu}(U) + m_\ell + i \mu_\ell \tau^3 \gamma^5 
    \right] \chi_\ell(x)\,,\nonumber\\
  S_{tm}^h =& \sum_x \bar{\chi}_h(x)\left[ D_W(U) + \frac{i}{4} \csw \sigma^{\mu\nu}
    \mathcal{F}^{\mu\nu}(U) + m_h - \mu_\delta \tau_1 + i \mu_\sigma \tau^3 \gamma^5
  \right] \chi_h(x)\,.
\end{align}
Here, $\chi_\ell=(u,d)^\top$ represents the light quark doublet and
$\chi_h=(s,c)^\top$ the heavy quark doublet. $\mu_\ell$ is the twisted
quark mass of the degenerate light doublet and $\mu_\delta$ and
$\mu_\sigma$ are the twisted quark masses of the heavy
doublet. $m_\ell$ and $m_h$ are the (untwisted) Wilson quark masses
and $D_W$ is the massless Wilson-Dirac operator. The bare untwisted
Wilson masses are tuned to the critical value $m_\ell = m_h =
m_\textrm{crit}$, which gives automatic $\mathcal{O}(a)$
improvement~\cite{Frezzotti:2000nk} and requires no further operator
level improvements. The clover term multiplied by the
Sheikoleslami-Wohlert improvement coefficient $\csw$  reduces
isospin symmetry breaking effects~\cite{PhysRevD.74.034501}, since
${\cal O}(a)$ improvement is already achieved from tuning to the critical
Wilson quark mass. The key parameters of the ensemble are collected in
Table~\ref{tab:params} and the remaining parameters are
$\kappa_\textrm{crit}=1/(2am_\textrm{crit}+8)=0.1400645$, $c_\mathrm{sw}=1.74$, $a\mu_{\ell}=0.003$,
$a\mu_\sigma=0.1408$, and $a\mu_\delta=0.1521$.

As in Ref.~\cite{Alexandrou:2020gxs} in which the lower two moments of
the Kaon PDF were computed, in the valence sector we use the so-called
Osterwalder Seiler fermions for the strange quarks, with the same
value of the bare strange quark mass.

\begin{table}[h!]
\centering
\renewcommand{\arraystretch}{1.2}
\renewcommand{\tabcolsep}{6pt}
\begin{tabular}{| l| c | c | c | c | c  | c | c |}
    \hline
    \multicolumn{8}{|c|}{Parameters} \\
    \hline
    Ensemble   & $\beta$ & $a$ [fm] & volume $L^3\times T$ & $N_f$ & $m_\pi$ [MeV] &
    $L m_\pi$ & $L$ [fm]\\
    \hline
    cA211.30.32 & 1.726 & 0.093  & $32^3\times 64$  & $u, d, s, c$ & 260
    & 4 & 3.0 \\
    \hline
\end{tabular}
\caption{Parameters of the ensemble used in this work.}
\label{tab:params}
\vspace*{0.2cm}
\end{table}

The matrix elements $\langle M(p)| \Op^{\{\mu\nu\rho\sigma \}} |M(p)\rangle$ use the interpolating fields of $\pi^+$ and $K^+$, that is, $J_{\pi^+} = \overline{d}\gamma_5 u$ and $J_{K^+} = \overline{s}\gamma_5 u$. For the pion, we only need to calculate the up quark contribution to $\avgxxx$, as the equality $G_{u}(x,x')=\gamma_5 G_{d}^\dagger(x',x)\gamma_5$ holds for twisted mass fermion propagators. The interpolating fields at the source and the sink are smeared using Gaussian smearing. Further details can be found in Ref.~\cite{Alexandrou:2020gxs}. 
The three-point correlation functions in the forward limit are given by
\begin{equation}
\label{eq:C3pt}
C_M^{1234}(t,t_s,\vec{p}) = \sum_{\vec{x}_s,\vec{x}} \langle 0 | J_M(t_s,\vec{x}_s){\cal O}^{1234}(t,\vec{x})J^\dag_M(t_i,\vec{x}_i) |  0 \rangle  e^{-i\vec{p}\cdot(\vec{x}_s-\vec{x}_i)}\,,
\end{equation} 
where $t_i$, $t$, $t_s$ denote the source, insertion and sink Euclidean times, respectively. Similarly, the spatial coordinates of the source, current insertion and sink are $\vec{x}_i$, $\vec{x}$, $\vec{x}_s$. We set the source to be at $t_i=0$, so that the source-sink separation is $t_s -t_i \equiv t_s$. In the results presented in this paper, we focus on the $u^+$ contribution to the pion, where $q^+\equiv q+\bar{q}$. To get the total connected contribution one may use the relation $\avgxxx_\pi^{u^+ + {d}^+}=2 \avgxxx_\pi^{u^+}$. 

\begin{table}[h!]
\centering
\renewcommand{\arraystretch}{1.2}
\renewcommand{\tabcolsep}{6pt}
\begin{tabular}{| c | c | c | c | c |}
    \hline
    \multicolumn{5}{|c|}{Statistics} \\
    \hline
    $t_s/a$ & $\#$ configurations   & $\#$ source positions & $\#$ momentum boost & Total statistics \\
    \hline
    \multicolumn{5}{|c|}{Two-point correlators} \\
    \hline
    --- & 122 & 72 & 8 & 70,272 \\
    \hline
    \multicolumn{5}{|c|}{Three-point correlators} \\
    \hline
    12 & 122 & 16 & 8 & 15,616 \\
    14, 16, 18 & 122 & 72 & 8 & 70,272 \\
    \hline
\end{tabular}
\caption{Statistics used in the calculation of $\avgxxx$.}
\label{tab:statistics}
\vspace*{0.2cm}
\end{table}

The calculation requires a large number of statistics to control the gauge noise introduced by the covariant derivatives. The signal-to-noise ratio is also suppressed due to the use of a boosted frame, which is necessary to obtain $\avgxxx$ directly in the forward kinematic limit while avoiding operator mixing. We use the smallest possible values, that is, momenta of the class $\vec{p}_i=\frac{2\pi}{L} (\pm1,\pm1,\pm1)$ with $\vec{p}_i\,^2=\frac{12\pi^2}{L^2}$. This leads to a factor of eight more computational cost, which is compensated by the reduction of statistical uncertainties by a factor of about $1/\sqrt{8}$. 

We analyze 122 configurations, separated by 20 trajectories to reduce auto-correlation effects. In order to control the gauge noise, we calculate the matrix elements for $\avgxxx$ for more than double the statistics as compared to our previous work on $\avgx$ and $\avgxx$~\cite{Alexandrou:2020gxs}. Based on the conclusions of Ref.~\cite{Alexandrou:2020gxs}, we use $t_s/a=14,\,16,\,18$ to reliably suppress excited states. The pion, being a lighter meson, suffers from higher statistical noise. Therefore, we add $t_s/a=12$ in the procedure for eliminating excited-states contamination. The statistics used for each value of $t_s$ are listed in Table~\ref{tab:statistics}.

\section{Renormalization\label{sec:III}}
\label{sec:renorm}

The three-derivative operator, in its general case, exhibits mixing with lower dimension operators~\cite{Capitani:1994qn,Beccarini:1995iv,Gockeler:1996mu,Gockeler:2004wp}. As mentioned in the previous section, we choose all indices of the operator to be different ($\mu\ne\nu\ne\rho\ne\sigma\ne\mu$), which avoids such a mixing and, therefore, its renormalization is multiplicative. We calculate the corresponding renormalization function, $Z_{\rm vDDD}$, non-perturbatively following the procedure we developed in Refs.~\cite{ Alexandrou:2010me,Alexandrou:2012mt,Alexandrou:2015sea}. We start by writing the bare vertex function as
\begin{equation}\label{vfun}  
   G(p) = \frac{a^{12}}{V}\sum_{x,y,z,z'} e^{-ip(x-y)} \langle u(x) \overline u(z) \J(z,z') d(z') \overline d(y) \rangle \, ,
\end{equation}
with $\J = \gamma^\mu D^\nu D^\rho D^\sigma$, and $u$ and $d$ representing quark fields in the physical basis. $p$ is the vertex momentum allowed by the boundary conditions, and $V$ is the lattice volume. The Dirac and color indices of $G(p)$ are suppressed for simplicity. We employ the momentum source approach, introduced in Ref.~\cite{Gockeler:1998ye}, which uses directly Eq.~(\ref{vfun}) with a source that is momentum dependent. For twisted mass fermions, we make use of the symmetry $S^u(x,y)=\gamma_5S^{d\dagger}(y,x)\gamma_5$ between the $u-$ and $d-$quark propagators, and therefore, extract the vertex function with a single inversion per momentum. While this method requires separate inversions for each momentum employed in the calculation, it has the advantage of high statistical accuracy and the evaluation of the vertex for any operator at no
significant additional computational cost. 

$Z_{\rm vDDD}$ is calculated by employing the Rome-Southampton method (RI$'$ scheme)~\cite{Martinelli:1994ty}, which involves the amputated vertex function
\begin{equation}
   \Gamma(p) = (S^u(p))^{-1}\, G(p)\, (S^d(p))^{-1} \, .
\label{vertexfunction}
\end{equation}
$S^u(p)$ and $S^d(p)$ are the propagators in momentum space in the physical basis, defined by
\begin{equation}\label{pprop}
   S^u(p) = \frac{a^8}{V}\sum_{x,y} e^{-ip(x-y)} \left\langle u(x) \overline u(y) \right\rangle\, , \qquad
   S^d(p) = \frac{a^8}{V}\sum_{x,y} e^{-ip(x-y)} \left\langle d(x) \overline d(y) \right\rangle \, .
\end{equation}
In practice, we work in the twisted basis at maximal twist, in which Eq.~(\ref{vfun}) takes the form
\begin{equation}\label{vfun_tm}
   G(p) = \frac{a^{12}}{4V}\sum_{x,y,z,z'} e^{-i p(x-y)} \left\langle(\eins+i\gamma_5) u(x) \overline u(z)(\eins+i\gamma_5) \J(z,z') (\eins-i\gamma_5) d(z')\
 \overline d(y)(\eins-i\gamma_5) \right\rangle \, .
\end{equation}
The above expression can be simplified by using $\s^u(x,z)=\gamma_5 {\s^d}^\dagger(z,x)\gamma_5$. The renormalization function in the RI$'$ scheme are determined by the conditions
\bea
\label{renormalization cond}
   Z_q^{-1}\,{\cal Z}_{\rm vDDD}\,\frac{1}{12} {\rm Tr} \left[\Gamma_{\rm vDDD}^L(p)
     \,\left(\Gamma_{\rm vDDD}^{{\rm Born}}(p)\right)^{-1}\right] \Bigr|_{p^2=\mu_0^2} = 1\, , \\[1.5ex]
   Z_q = \frac{1}{12} {\rm Tr} \left[(S^L(p))^{-1}\, S^{{\rm Born}}(p)\right] \Bigr|_{p^2=\mu_0^2}\,, \qquad\quad
\eea
where the trace is taken over spin and color. The momentum of the vertex function is indicated by $p$, and is set to the RI$'$ renormalization scale, $\mu_0$. $S^{{\rm Born}}$ ($\Gamma_{\rm vDDD}^{{\rm Born}}$) is the tree-level value of the fermion propagator (operator). We differentiate ${\cal Z}_{\rm vDDD}$ and $Z_{\rm vDDD}$, as the former depends on the pion mass of the ensemble and the initial RI$'$ renormalization scale. The latter is our final estimate after the chiral extrapolation and after the limit $(a\mu_0)^2\to 0$ has been taken.

We evaluate the vertex functions and propagators for a wide range of values for $(a\,p)^2$, using 10 gauge configurations, which leads to per mil statistical accuracy. We use momenta that have the same spatial components, that is:
\begin{equation}
\label{eq:P4}
(a p) \equiv 2\pi \left(\frac{n_t}{L_t}+\frac{1}{2\,L_t},
\frac{n_x}{L_s},\frac{n_x}{L_s},\frac{n_x}{L_s}\right)\,,  \qquad\quad n_t \,\epsilon\, [2, 9]\,,\quad n_x\,\epsilon\, [2, 5]\,,\quad (a p)^2 \in [0.9 - 6.7]\,,
\end{equation} 
where $L_t$ ($L_s$) is the temporal (spatial) extent of the lattice. Democratic momenta in the spatial directions that satisfy, in addition, Eq.~(\ref{eq:P4})  reduce non-Lorentz invariant contributions ($ {\sum_i p_i^4}/{(\sum_i p_i^2 )^2}{<}0.3$). This is based on empirical arguments~\cite{Constantinou:2010gr} and is being implemented in all calculations by our group. We further improve ${\cal Z}_{\rm vDDD}$ by subtracting the ${\cal O}(g^2\,a^\infty)$ artifacts from $Z_q$, which enters the renormalization condition of Eq.~(\ref{renormalization cond}). The artifacts are computed to one loop in perturbation theory and to all orders in the lattice spacing, ${\cal O}(g^2\,a^\infty)$, as outlined in Refs.~\cite{Constantinou:2014fka,Alexandrou:2015sea}. Note that the vertex function of the three-derivative operator also contain ${\cal O}(g^2\,a^\infty)$, but have not been calculated yet.

We obtain $Z_{\rm vDDD}$ on five $N_f=4$ ensembles at the same lattice spacing as the $N_f=2+1+1$ ensemble we use for the meson matrix elements. The ensembles correspond to different pion masses, and are used in order to take the chiral limit. The parameters of the $N_f=4$ ensembles used for $Z_{\rm vDDD}$ are given in Table~\ref{tab:Z_ensembles}. The chiral limit is taken using a quadratic fit with respect to the pion mass, or linear in $a\mu$, giving a zero slope within uncertainties in both cases.
\begin{table}[!h]
\begin{center}
\begin{tabular}{ccc}
\hline
\hline
$\,\,\,$        $\,\,\,$              & $\beta=1.726$, $\,\,\,a=0.093$ fm  &          \\ \hline
\hline
$a \mu$   & $a m_{PS}$ & lattice size\\
\hline
$\,\,\,$  0.0060$\,\,\,$         & 0.1680   & $24^3 \times 48$ \\  
$\,\,\,$  0.0080$\,\,\,$         & 0.1916   & $24^3 \times 48$ \\          
$\,\,\,$  0.0100$\,\,\,$         & 0.2129   & $24^3 \times 48$ \\  
$\,\,\,$  0.0115 $\,\,\,$        & 0.2293   & $24^3 \times 48$ \\ 
$\,\,\,$  0.0130 $\,\,\,$        & 0.2432   & $24^3 \times 48$  \\ 
\hline
\hline
\end{tabular}
\caption{Parameters for the $N_f=4$ ensembles used for the renormalization functions.} 
\label{tab:Z_ensembles}
\end{center}
\end{table}

\begin{figure}[!h]
\centerline{\includegraphics[scale=0.43]{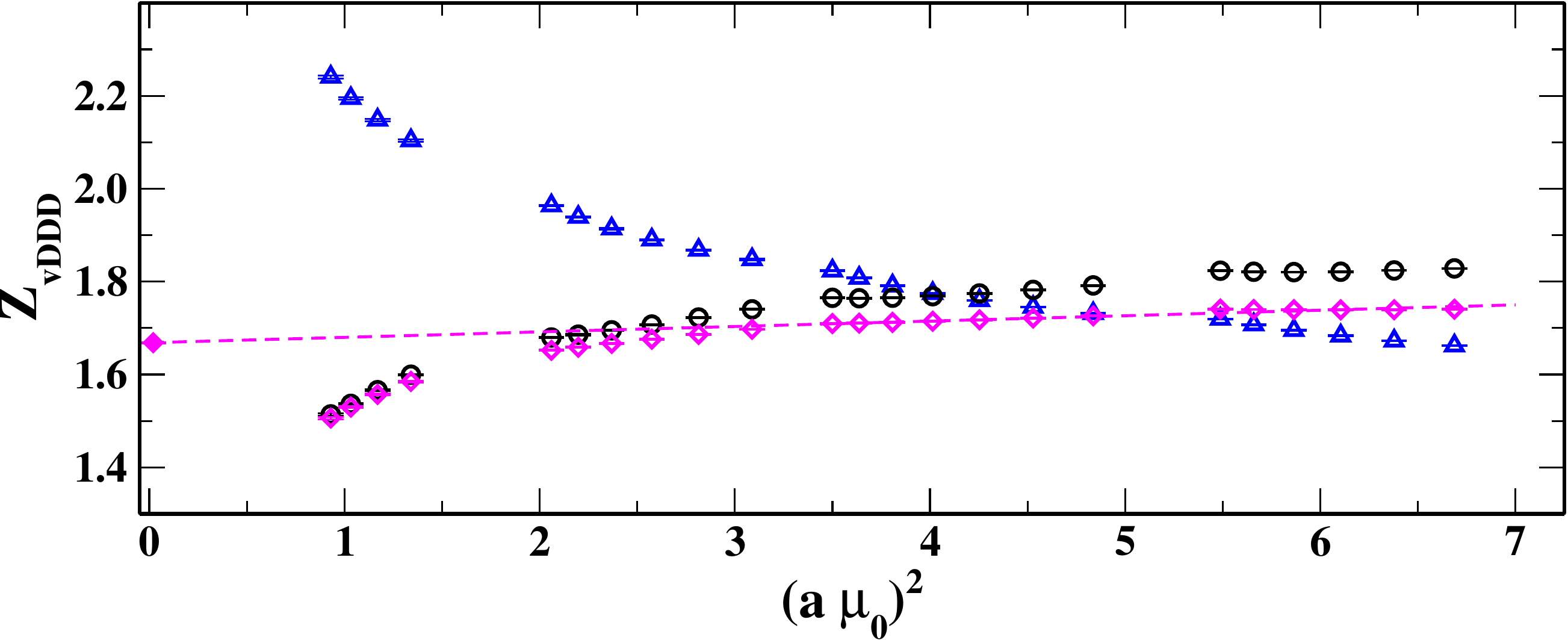}}
\caption{Chirally extrapolated $Z_{\rm vDDD}$ in the RI$'$ scheme (blue triangles), ${\overline{\rm MS}}$ scheme (black circles, magenta diamonds) as a function of the initial scale. The magenta diamonds correspond to the improved estimates upon subtraction of the artifacts in $Z_q$. The dashed line corresponds to the fit of Eq.~(\ref{Zfinal}). The filled magenta diamond is our final value for $Z_{\rm vDDD}$.}
\label{ZRIMS}
\end{figure}

The chirally extrapolated values for ${\cal Z}_{\rm vDDD}$ in the RI$'$ scheme are converted and evolved to $\overline{\rm MS} (2~{\rm GeV})$ using an intermediate Renormalization Group Invariant (RGI) scheme. Finally, a linear fit with respect to $(a \mu_0)^2$ is applied to the $\overline{\rm MS}$ estimates to eliminate residual dependence on $\mu_0$, that is
\begin{equation}
{\cal Z}_{\rm vDDD}(a\,p)= Z_{\rm vDDD} + Z_{\rm vDDD}^{(1)}\cdot(a\,\mu_0)^2\,.
\label{Zfinal}
\end{equation}
$Z_{\cal O}$ corresponds to the final value of the renormalization function for operator ${\cal O}$. The estimates for $Z_{\rm vDDD}$ in the RI$'$ and ${\overline{\rm MS}}$ schemes as a function of the initial RI$'$ renormalization scale are shown in Fig.~\ref{ZRIMS}. 
\noindent
$Z_{\rm vDDD}^{\overline{\rm MS}}$ is given at a scale of $\mu=2$ GeV. We find that, for $(a\mu_0)^2\ge 3$ the purely non-perturbative data exhibit a small residual dependence on the initial scale $\mu_0$ they were evolved from. The subtraction of the lattice artifacts in $Z_q$ results in a smaller slope, demonstrating the effectiveness of the artifact-subtraction procedure. We obtain 
\begin{equation}
{\cal Z}^{\overline{\rm MS}}_{\rm vDDD} (2\, {\rm GeV}) = 1.668(1)(26)\,,
\end{equation}
with the number in the first (second) parenthesis being the statistical (systematic) uncertainty. The indicated systematic effect is obtained by taking the deference between the estimates for the fit interval $(a\,\mu_0)^2\, \epsilon\, [2-7]$ and $(a\,\mu_0)^2\, \epsilon\, [3-7]$.

\section{Analysis methods\label{sec:analysis}}

\subsection{Setup}

To extract the meson matrix elements, one requires to take the ratio of the three-point correlation functions of Eq.~(\ref{eq:C3pt}) with the two-point functions, given by
\begin{equation}
    C_M(t,\mathbf{p})=\sum_{\mathbf{x}} \langle 0 | J_M(t,\mathbf{x})J_M^\dag (0,\mathbf{0})  | 0 \rangle e^{i\mathbf{p}\cdot\mathbf{x}}\,.
\end{equation}
The normalization of the meson state is $\langle 0 |J_M|M\rangle = Z_M/\sqrt{2 E}$. The ground state contribution can be isolated from the ratio
\begin{equation}
\label{eq:ratio}
 {\cal R}_M^{1234}  \equiv \frac{C^{1234}_M(t_s,t;\mathbf{p})}{C_M(t_s, \mathbf{p})}\,,
\end{equation}
which cancels unknown overlap terms between the interpolating field and the meson state. In this work, we apply two methods to obtain the ground state, namely a single-state (plateau) fit, and a two-state fit.

\subsubsection{Two-point function}

In both methods for identifying and eliminating excited states, analyse  the two-point function and perform  a single-state and two-state fit, the latter being
\begin{equation}
\label{C2pt_fit}
    C_M(\mathbf{p}^2; t) = c_0 \left( e^{-E_0(\mathbf{p}^2)t} + e^{-E_0(\mathbf{p}^2)(T - t)} \right) + c_1 \left( e^{-E_1(\mathbf{p}^2)t} + e^{-E_1(\mathbf{p}^2)(T - t)} \right) \,.
\end{equation}
In the above equation, the fitting parameters are the ground state energy, $E_0(\mathbf{p}^2)$, the first excited state energy, $E_1(\mathbf{p}^2)$, and the amplitudes $c_0$ and $c_1$. The plateau fit takes into account only the first term of Eq.~(\ref{C2pt_fit}). We fit the two-point functions averaged over the eight different directions of meson momentum ($\mathbf{p}^2 = 12 \pi^2 / L^2$), as the energies only depend on the momentum squared. In addition, using the averaged two-point functions improves the stability of the fit compared to fits on the individual momentum directions. The effective energy is calculated using the formula 
\begin{equation}
  \label{eq:effener}
    E_{\rm eff}(\mathbf{p}^2; t) = \frac{1}{2} \ln \left[ \frac{C_M(\mathbf{p}^2; t-1) + \sqrt{C_M(\mathbf{p}^2; t-1)^2 - C_M(\mathbf{p}^2; T/2)^2}}{C_M(\mathbf{p}^2; t+1) + \sqrt{C_M(\mathbf{p}^2; t+1)^2 - C_M(\mathbf{p}^2; T/2)^2}} \right] \,,
\end{equation}
which assumes that the two-point functions are symmetrized, i.e., that the two-point functions at $t$ have been averaged with the values at $T-t$. For the fits, we select a range of $t \in [t_{\rm low}, 31]$ with varying $t_{\rm low}$. We choose the lowest values of $t_{\rm low}$, such that the resulting ground state energies, $E_{\rm 2-state}$ and $E_{\rm plat}$ satisfy the condition
\begin{equation}
\label{eq:criterion}
    \frac{1}{2}\delta E_{\rm plat} \geq | E_{\rm plat} - E_{\rm 2-state} |\, ,
\end{equation}
where $\delta E_{\rm plat}$ is the error on the plateau fit. This criterion, while not unique, works very well on the lattice data we obtain here (see Fig.~\ref{fig:mEff_tlow_test_pion} and Fig.~\ref{fig:mEff_tlow_test_kaon}).

\subsubsection{Single-state fit}

We calculate the ratios in Eq.~(\ref{eq:ratio}) using a fit of the two-point function instead of the actual lattice data, that is
\begin{equation}
C_M(t) = c_0 \exp(-E_0 t)\,,
\end{equation}
where $c_0$ and $E_0$ are the ground state amplitude and energy, respectively. Using the modified two-point function removes $t_s$ dependence from the plateau values of the the ratios so that the plateaus converge at high $t_s$. Based on the plateau method, at insertion times far enough from the source and sink, the above ratio becomes time independent, i.e.,
\begin{equation}
    \frac{C_M^{1234}(t_s,t;\mathbf{p})}{C_M(t_s,\mathbf{p})} \xlongrightarrow[\text{$Et/a\gg1$}]{\text{$\Delta E(t_s-t)/a\gg1$}}  \Pi^{1234}_M(\mathbf{p})\,.
    \label{eq:threep_twop_ratio}
\end{equation}
In practice, we apply a constant fit in a region where a plateau is identified. The time-independent ratio (plateau) is renormalized multiplicative with $\mathcal{Z}_{\rm vDDD}$, and is related to the desired $\avgxxx$ as given in Eq.~(\ref{decomp2})
\begin{equation}
        \avgxxx_M = \frac{i\,\mathcal{Z}_{\rm vDDD} \, \Pi^{1234}_M(\mathbf{p})}{{p^1} {p^2} {p^3}}  \,.
        \label{eq:x3_plateau}
\end{equation}
Indeed, we confirm that the signal is found in the imaginary part of the three-point function. The single-state fit is applied on each $t_s$ separately, and the ground state is extracted at the lowest $t_s$, beyond which the plateau value is $t_s$-independent. We also compare with the two-state fit to confirm convergence.

\subsubsection{Two-state fit}

For the two-state fit we include the data for all $t_s$ simultaneously. The two-state fit is calculated by fitting the three-point functions to the Ansatz
\begin{equation}
    C_M^{1234}(t,t_s) = \:
    A_{00} e^{-E_0t_s} + A_{01} e^{-E_0(t_s-t)-E_1t}
    + A_{10} e^{-E_1(t_s-t)-E_0t}
    + A_{11} e^{-E_1t_s} \,,
\end{equation}
where the fitting parameters are the amplitudes $A_{00}$, $A_{10}$, and $A_{11}$ since, at zero momentum transfer, $A_{10}=A_{01}$. To avoid heavy notation we do not include a subscript $M$ in the parameters and energies. We use the energies $E_0$ and $E_1$ calculated from the two-state fit on the two-point functions. The results of the two-state fits on the two- and three-point functions are related to the matrix elements by
\begin{equation}
    \Pi^{1234}_M = \frac{A_{00}}{c_0} \,,
\end{equation}
so that $\avgxxx_M$ is calculated from the two-state fit as
\begin{equation}
    \avgxxx_M = \frac{i \,\mathcal{Z}_{\rm vDDD} \, A_{00}}{p_1 p_2 p_3 \, c_0} \,.
    \label{eq:x3_tsf}
\end{equation}
As mentioned above, the two-state fit is useful to confirm ground-state dominance beyond a certain value of $t_s$.

\subsection{Pion}
\label{sec:excited_states_pi}

\begin{figure}[h]
    \centering
    \includegraphics[scale=0.4]{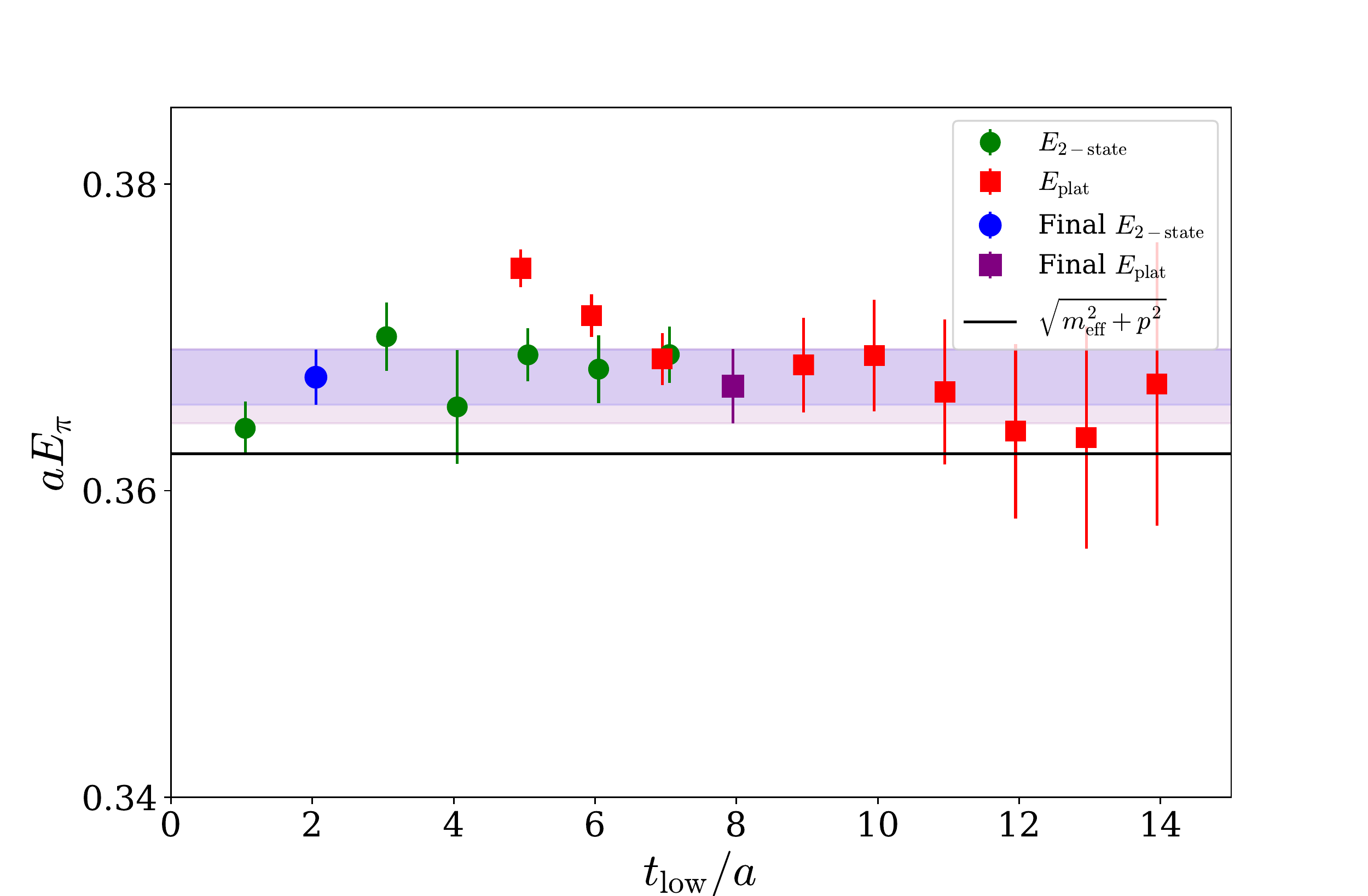}
\vspace*{-0.2cm}
    \caption{The pion ground state energies calculated using plateau fits (red squares) and two-state fits (green circles) for the range $t \in [t_{\rm low}, 31\,a]$. The black line shows the value from the continuum dispersion relation as explained in the text.}
    \label{fig:mEff_tlow_test_pion}
\end{figure}

We start our presentation with the extraction of the ground state energy for a boosted pion, which is needed for the analysis of the three-point functions. We follow the setup outlined above, and the results are shown in Fig.~\ref{fig:mEff_tlow_test_pion}. We apply a single-state (plateau), as well as two-state fits, and test the results against energies obtained via the continuum dispersion relation $E^2(\mathbf{p}) = m^2 + (\frac{2\pi}{L}\mathbf{p})^2$, where $m$ is obtained from the effective mass of the two-point correlation function at zero momentum, i.e. Eq.~(\ref{eq:effener}). For the two-state fit we vary $t_{\rm low}$ between $t=1a$ and $t=7a$. The single-state fit is applied for $t_{\rm low} \in [5 a - 14 a ]$. As can be seen in Fig.~\ref{fig:mEff_tlow_test_pion}, we find that the ground-state energy is isolated already at $t_{\rm low}/a = 2$ for the two-state fit and at $t_{\rm low}/a = 8$ for the plateau fit. The ground-state energy obtained from the plateau fit is $aE_{\rm plat}= 0.3668(24)$ and from the two-state fit is $aE_{2-{\rm state}}= 0.3674(18)$. These values satisfy the criterion of Eq.~(\ref{eq:criterion}), and also, are compatible with each other and with the dispersion relation.

\begin{figure}[h]
    \centering
    \includegraphics[scale=0.475]{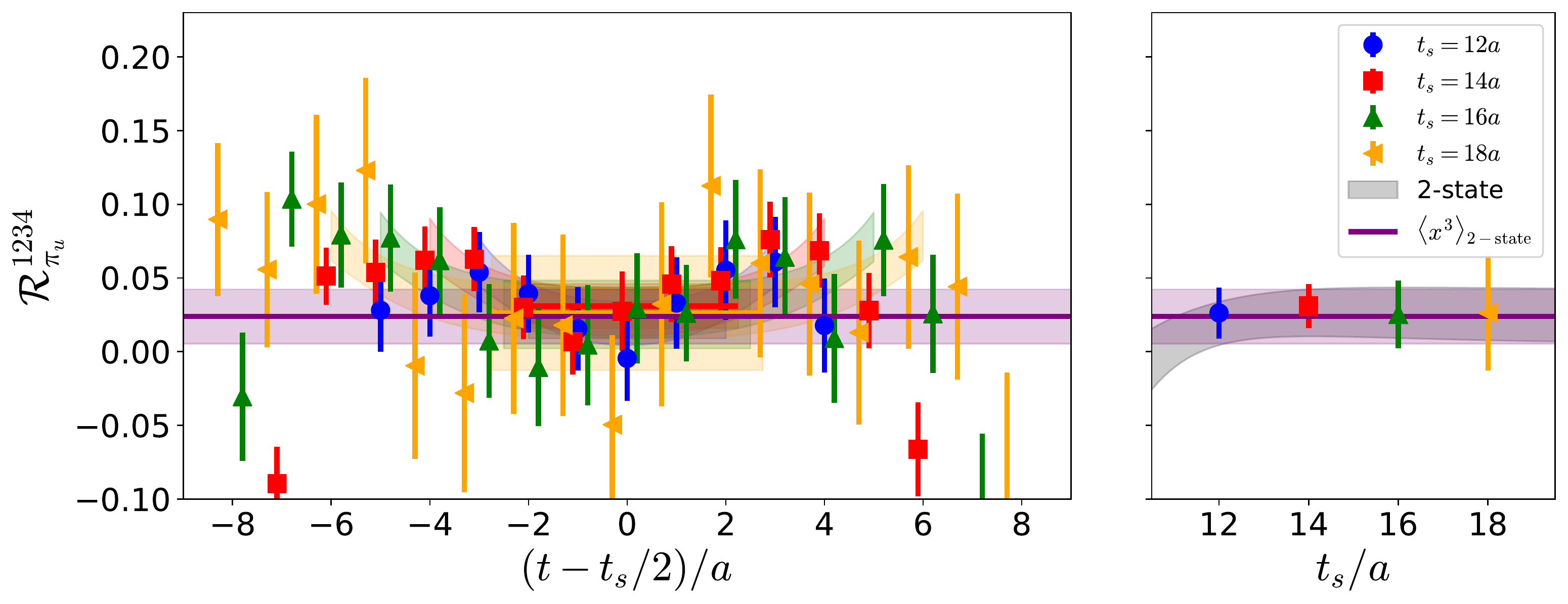}
    \vspace*{-0.2cm}
    \caption{Left panel: The ratios leading to $\avgxxx^u_\pi$ for the pion using Eq.~(\ref{eq:x3_plateau}) for $t_s/a=12,\, 14,\, 16,\, 18$, shown with blue circles, red squares, green up triangles and orange left triangles, respectively. The plateau fits for each $t_s$ are shown with bands of the same color as the data points. The purple line in both panels is the value of $\avgxxx^u_\pi$ calculated from the two-state fit according to Eq.~(\ref{eq:x3_tsf}). Right panel: The $\avgxxx^u_\pi$ values extracted from the plateau fit of the left panel (blue, red, green, orange points). The gray curve is the ratio of Eq.~(\ref{eq:threep_twop_ratio}) as obtained using the two-state fit parameters for $t = t_s/2$.}
    \label{fig:x3_pion}
\end{figure}

Using the above values for the ground state energy we form the modified two-point function of Eq.~(\ref{C2pt_fit}) and take the ratio of Eq.~(\ref{eq:ratio}). We also ensure that excited-state effects can be successfully suppressed by calculating the three-point correlators at different source-sink time separations. In our previous study~\cite{Alexandrou:2020gxs}, we calculated $\avgx$ in the rest frame for $t_s / a=12,\, 14,\, 16,\, 18,\, 20,\, 24$ and found that the results converge at $t_s/a\geq18$. Therefore, for the boosted frame we focus on $t_s / a=12,\,14,\,16,\,18$. Fig.~\ref{fig:x3_pion} shows the ratios which lead to $\avgxxx_\pi^u$, that is, the right-hand-side (rhs) of Eq.~(\ref{eq:x3_plateau}) for the plateau method and the rhs of Eq.~(\ref{eq:x3_tsf}) for the two-state fit. We find that the signal is more noisy for $t_s / a \geq 16$, due to the three covariant derivatives and the boosted frame setup. All plateau fits, however, are found to be consistent with one another as well as with the two-state fit result. This finding suggests that any remaining excited-state effects are within the reported errors. The plateau and two-state fit values are collected in Table~\ref{table:x3_pi}. For completeness, we give the updated values for $\avgxx_\pi^u$ obtained with double the statistics for $t_s / a=14,\,16,\,18$ compared to our previous work~\cite{Alexandrou:2020gxs}. The computational challenges associated with the gauge noise contamination in three-derivative operators, the boosted frame with three nonzero spatial components, coupled with the light mass of the pion are reflected in the increased uncertainties in $\avgxxx_\pi^u$. However, this is not the case for the kaon, which is about twice heavier than the pion for this ensemble (see Table~\ref{table:avgX3kaon}).

\begin{table}[h!]
    \begin{tabular}{ccc}
      \hline\hline\\[-2.5ex]
 $t_s/a$ & $\,\,\avgxx^u_\pi$ & $\,\,\avgxxx^u_{\pi}$ \\[0.5ex]
\hline\\[-2.5ex]
$\,\,\,$12  &$\,\,\,$0.110(6) & $\,\,\,$0.026(17)      \\
$\,\,\,$14  &$\,\,\,$0.114(5) & $\,\,\,$0.031(15)      \\
$\,\,\,$16  &$\,\,\,$0.105(9) & $\,\,\,$0.025(23)      \\
$\,\,\,$18  &$\,\,\,$0.099(15) & $\,\,\,$0.026(39)      \\
2-state     &$\,\,\,$0.110(7) & $\,\,\,$0.024(18)      \\ [0.5ex]
      \hline\hline
    \end{tabular}
        \caption{Renormalized values for the pion $\avgxx$ and $\avgxxx$ using a single-state fit on $t_s/a$=12, 14, 16, 18, and the corresponding two-state fit. The results are given in the $\overline{\rm MS}$ scheme at 2 GeV. The number shown in the parenthesis is the statistical error.}
        \label{table:x3_pi}
  \end{table}

\subsection{Kaon}

\begin{figure}[h]
    \centering
    \includegraphics[scale=0.4]{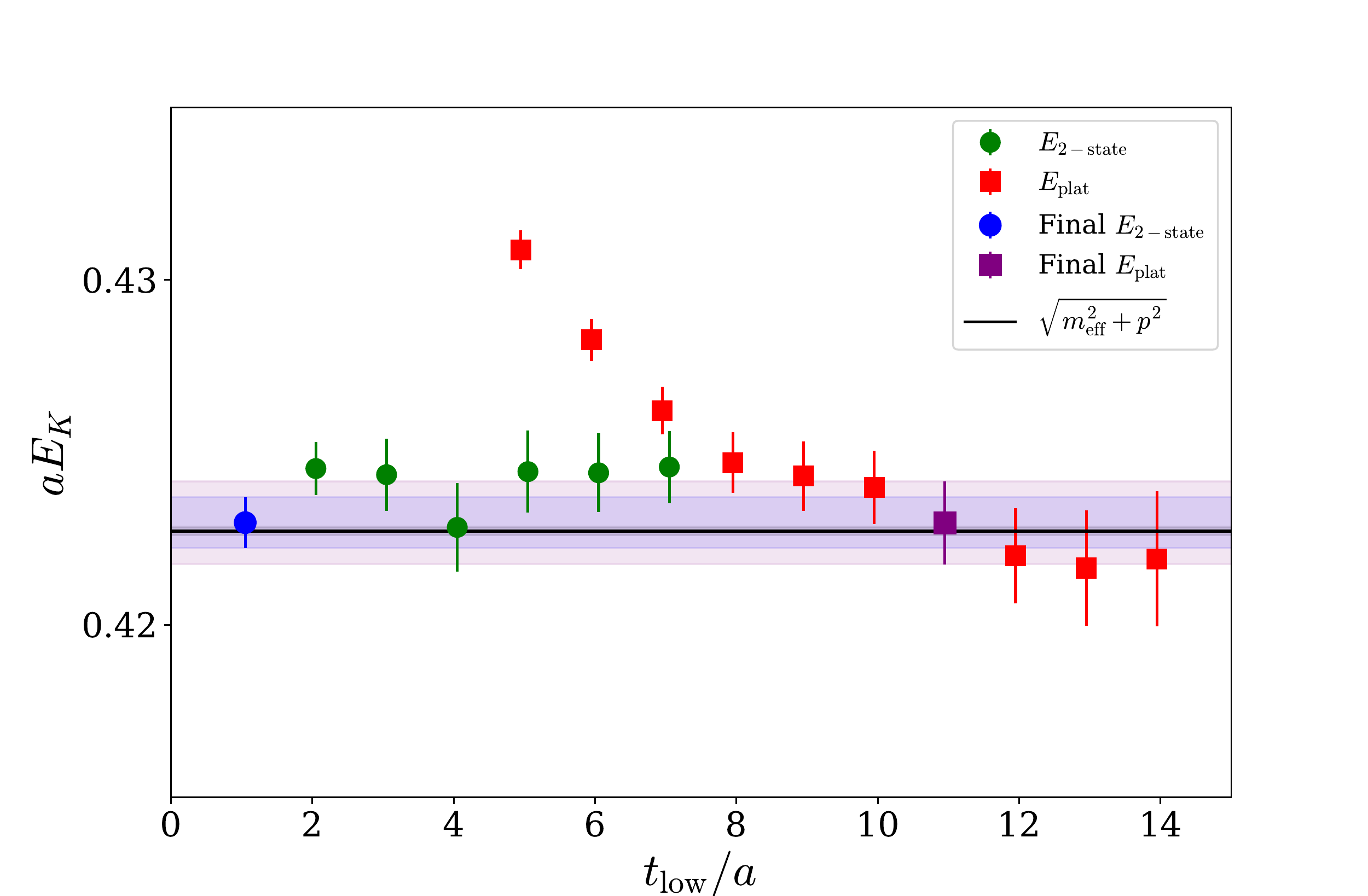}
    \caption{The kaon ground state energies calculated by both fits using $t \in [t_{\rm low}, 31\,a]$. The notation is the same as in Fig.~\ref{fig:mEff_tlow_test_pion}.}
    \label{fig:mEff_tlow_test_kaon}
\end{figure}

Our analysis for the kaon follows the same procedure for the two-point and three-point correlation functions, as in the case of the pion. In Fig.~\ref{fig:mEff_tlow_test_kaon}, we show the extraction of the ground-state energy, $E_K$. The dependence of $E_K$ on $t_{\rm low}$ is similar to the pion. We extract the plateau value $aE_{\rm plat}=0.4230(12)$ from $t_{\rm low}=11 a$, and the two-state fit value $aE_{2-{\rm state}}= 0.4230(7)$ at $t_{\rm low}=1a$. 

\begin{figure}[h]
    \centering
    \includegraphics[scale=0.45]{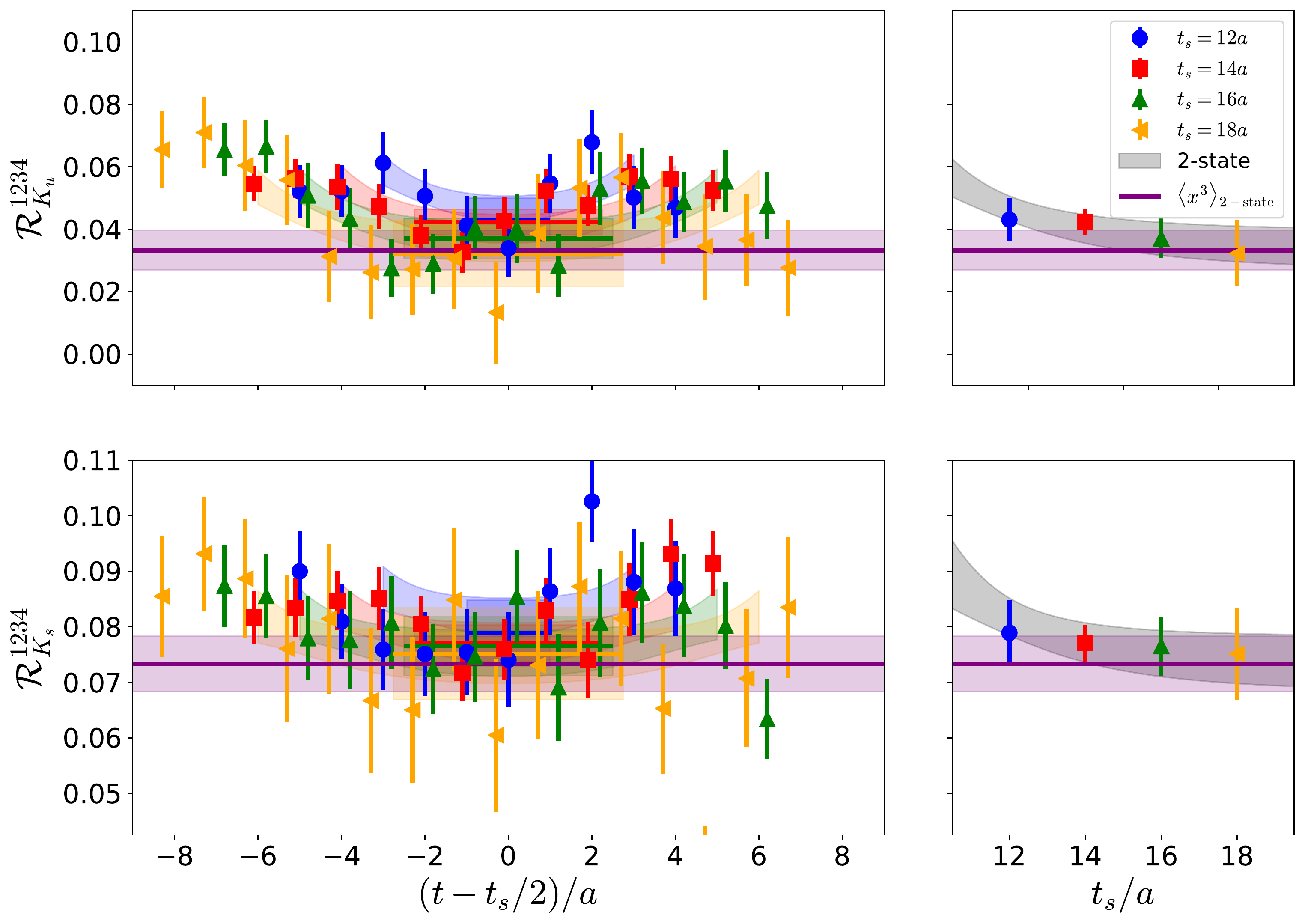}
    \caption{The ratios leading to $\avgxxx$ for the kaon. The top and bottom panels correspond to the up and strange contributions, respectively. The notation is the same as in Fig.~\ref{fig:x3_pion}.}
    \label{fig:x3_kaon}
\end{figure}

In Fig.~\ref{fig:x3_kaon}, we plot the ratio leading to $\avgxxx_K$ for the up and strange quark. We show the four values of $t_s$ and compare with the two-state fit. As can be seen, the gauge noise is decreased compared to the pion, attributed to the heavier mass of the kaon. There is a clear signal for both flavors of the kaon for all source-sink time separations. Similarly to the pion, the plateau values are consistent with the two-state fit. The plateau and two-state fits are collected in Table~\ref{table:avgX3kaon}.
\begin{table}[h!]
    \begin{tabular}{ccccc}
      \hline\hline\\[-2.5ex]
 $t_s/a$ & $\,\,\avgxx^u_{K}$ & $\,\,\avgxx^s_{K}$  & $\,\,\avgxxx^u_{K}$ & $\,\,\avgxxx^s_{K}$ \\[0.5ex]
\hline\\[-2.5ex]
$\,\,\,$12  &$\,\,\,$0.101(2)   &$\,\,\,$0.146(2) &$\,\,\,$0.043(7)  &$\,\,\,$00.079(6) \\
$\,\,\,$14  &$\,\,\,$0.099(2)   &$\,\,\,$0.142(2) &$\,\,\,$0.042(4)   &$\,\,\,$0.077(3) \\
$\,\,\,$16  &$\,\,\,$0.096(2)   &$\,\,\,$0.139(2)  &$\,\,\,$0.037(6)   &$\,\,\,$0.077(5) \\
$\,\,\,$18  &$\,\,\,$0.095(3)   &$\,\,\,$0.138(3)   &$\,\,\,$0.032(11)   &$\,\,\,$0.075(8) \\
2-state     &$\,\,\,$0.096(2)   &$\,\,\,$0.139(2)    &$\,\,\,$0.033(6)   &$\,\,\,$0.073(5)     \\ [0.5ex]
      \hline\hline
    \end{tabular}
        \caption{Renormalized values for the up- and strange-quark contribution to the kaon $\avgxx$ and $\avgxxx$, for the single-state fits and the corresponding two-state fit.        The results are given in the $\overline{\rm MS}$ scheme at 2 GeV. The number shown in the parenthesis is the statistical error.}
        \label{table:avgX3kaon}
  \end{table}

\section{Results - comparison with other studies}
\label{sec:fin_results}

\subsection{Results\label{sec:IV}}

Based on our analysis, we choose the results extracted using two-state fits as our final values for all quantities. We report as systematic uncertainties the difference between the  value extracted using the two-state fit and that determined using the plateau fit at $t_s=18 a$. Our final results in the $\overline{\rm MS}$ at a scale of 2 GeV are 
\bea
\label{eq:x2_p_u}
\avgxx^u_{\pi^+} & = & 0.110(7)(12)\,, \\[1.5ex]
\label{eq:x2_k_u}
\avgxx^u_{K^+} & = & 0.096(2)(2)\,, \\[1.5ex]
\label{eq:x2_k_s}
\avgxx^s_{K^+} & = & 0.139(2)(1)\,,
\eea
for the second moment, and 
\bea
\label{eq:x3_p_u}
\avgxxx^u_{\pi^+} & = & 0.024(18)(2)\,, \\[1.5ex]
\label{eq:x3_k_u}
\avgxxx^u_{K^+} & = & 0.033(6)(1)\,, \\[1.5ex]
\label{eq:x3_k_s}
\avgxxx^s_{K^+} & = & 0.073(5)(2)\,,
\eea
for the third. We also calculate the ratios $\avgxxx/\avgx$ and $\avgxxx/\avgxx$. The calculation of $\avgx$ and $\avgxx$ is discussed in Ref.~\cite{Alexandrou:2020gxs}. These ratios are of interest because they reveal partial information on the $x$-dependence PDFs, in particular, the large-$x$ region. In general, the statistical errors for $\avgxxx$ are larger than the lower moments, which is propagated to the ratios. The $\avgxxx$ to $\avgx$ ratios are
\bea
\label{eq:x_over_x3_pi_fin}
\frac{\avgxxx_\pi^{u^+}}{\avgx_\pi^{u^+}} &=& 0.092(71)(6)\,,\\[1.5ex]
\label{eq:x_over_x3_k_u_fin}
\frac{\avgxxx_K^{u^+}}{\avgx_K^{u^+}} &=& 0.135(26)(8)\,,\\[1.5ex]
\label{eq:x_over_x3_k_s_fin}
\frac{\avgxxx_K^{s^+}}{\avgx_K^{s^+}} &=& 0.232(16)(1)\,,
\eea
and the $\avgxxx$ to $\avgxx$ ratios are
\bea
\label{eq:x2_over_x3_pi_fin}
\frac{\avgxxx_\pi^{u^+}}{\avgxx_\pi^{u^+}} &=& 0.216(167)(50)\,,
\label{eq:x2_over_x3_k_u_fin} \\[1.5ex]
\frac{\avgxxx_K^{u^+}}{\avgxx_K^{u^+}} &=& 0.346(66)(5)\,,\\[1.5ex]
\label{eq:x2_over_x3_k_s_fin}
\frac{\avgxxx_K^{s^+}}{\avgxx_K^{s^+}} &=& 0.529(36)(18)\,.
\eea
It is also interesting to compare $\langle x^n \rangle_\pi$ and $\langle x^n \rangle_K$, as it is related to the SU(3) flavor symmetry breaking due to the heavier mass of the strange quark. This is important because some quantities may be sensitive to SU(3) flavor symmetry breaking, such as the pion and kaon radii~\cite{Zyla:2020zbs}. For the comparison of the up quark contributions we find
\bea
\label{eq:x_K_over_x_pi_fin}
\frac{\avgx_\pi^{u^+}}{\avgx_K^{u^+}} &=& 1.060(9)(7)\,, \\[1.5ex]
\label{eq:x2_K_over_x2_pi_fin}
\frac{\avgxx_\pi^{u^+}}{\avgxx_K^{u^+}} &=& 1.148(57)(106)\,, \\[1.5ex]
\label{eq:x3_K_over_x3_pi_fin}
\frac{\avgxxx_\pi^{u^+}}{\avgxxx_K^{u^+}} &=& 0.717(488)(94) \,,
\eea
while for the strange quark in the kaon over the up quark in the pion we have
\bea
\frac{\avgx_\pi^{u^+}}{\avgx_K^{s^+}} &=& 0.823(8)(10)\,, \\[1.5ex]
\frac{\avgxx_\pi^{u^+}}{\avgxx_K^{s^+}} &=& 0.795(45)(80)\,, \\[1.5ex]
\frac{\avgxxx_\pi^{u^+}}{\avgxxx_K^{s^+}} &=&  0.325(244)(23) \,.
\eea
We find that the SU(3) symmetry breaking is $\sim 5-10\%$ for $\avgx$ and $\sim 10-20\%$ for $\avgxx$. The results for $\avgxxx$ indicate a symmetry breaking of $\sim 30-50\%$, with larger uncertainties. These results are very interesting because based on intuitive arguments, the strange quark PDF has its support at higher $x$ than the up quark PDF, which indicates that the symmetry breaking is more pronounced in the higher moments. 
We will discuss further the SU(3) flavor symmetry breaking in Section~\ref{sec:PDFs}. 

\subsection{Other lattice calculations}

There are only a few direct calculations for $\avgxxx$ extracted from the three-derivative local vector operator, starting with the pioneering work of QCDSF-UKQCD in 1997~\cite{Best:1997qp}, which was later extended in 2007~\cite{Brommel:2006zz,Brommel:2007zz}, and reanalyzed in Ref.~\cite{Detmold:2003tm}. All these calculations are for the pion; our results are the first reported for the kaon, using the three-derivative local operator. Interest in the pion and kaon structure has recently been renewed, due to novel approaches to extract the $x$-dependence of PDFs, such as the quasi-PDFs~\cite{Ji:2013dva,Ji:2014gla}, the pseudo-Ioffe-time-distributions (ITD)~\cite{Radyushkin:2016hsy}, and current-current correlators~\cite{Ma:2014jla,Ma:2014jga,Ma:2017pxb}. Using such methods, the lowest moments of the pion~\cite{Joo:2019bzr,Lin:2020ssv,Gao:2020ito}, and the kaon~\cite{Lin:2020ssv} have been obtained~\cite{Joo:2019bzr,Lin:2020ssv,Gao:2020ito}, either from integration on the $x$-dependent PDF, or via the so-called operator product expansion without an operator product expansion (OPE without OPE) method~\cite{Martinelli:1998hz}. Given the small number of calculations, we compare our results with all these methods. One has to bear in mind that, each calculation has its own systematic uncertainties and uses a different methodology. Therefore, the comparison is qualitative at this stage, as not all sources of systematic uncertainties have been quantified.

The calculation of Ref.~\cite{Best:1997qp} was done in the quenched approximation for Wilson fermions using ensembles with pion mass 712, 1013, 1208 MeV. They used an operator with only two indices different (${\cal O}_{v4}= {\cal O}^{ii44}$), which requires momentum boost only in one spatial direction. However, this operator exhibits mixing with lower dimensional operators, which is difficult to eliminate. The calculation ignored the mixing and $\avgxxx_\pi$ was renormalized using results from perturbation theory. The reported values for $\avgxxx^u_\pi$ in the $\overline{\rm MS}$ scheme at a scale of 2.4 GeV are 0.0619(45) 0.0580(65) 0.054(18) for the ensembles with pion mass 1208, 1013, 712 MeV, respectively. In Table~\ref{tab:x3_pi_compare_lattice} we present their estimate for the extrapolated value, 0.048(20), after evolution to 2 GeV. The results of Ref.~\cite{Best:1997qp} have been analyzed in Ref.~\cite{Detmold:2003tm} using different methods to perform the chiral extrapolation on the three ensembles mentioned above. For two of the methods they report $\avgxxx^u_\pi$= 0.043(15)(3), and 0.05(2) at a scale of 2.4 GeV. The value of the first method is evolved to 2 GeV and is given in Table~\ref{tab:x3_pi_compare_lattice}.

Almost a decade after their first calculation, QCDSF-UKQCD has improved their work in more than one ways, as presented in Refs.~\cite{Brommel:2006zz}. The ensembles employed are unquenched ($N_f=2$) clover fermions. Several ensembles were used to extract $\avgxxx_\pi$ with pion mass between 450 - 1180 MeV. The operator ${\cal O}_{v4}$ was used, and while the mixing was ignored, the renormalization was done non-perturbatively. Their preliminary results chirally extrapolated to the physical pion mass gave $\langle x^3\rangle_\pi$ = 0.074(9)(4) in the $\overline{\rm MS}$ at a scale of 2 GeV. It should be noted, that the accuracy of the chirally extrapolated value is heavily influenced  from the accuracy of the ensembles with $m_\pi=800$ MeV and higher. Another update of the calculation was presented in the thesis of Ref.~\cite{Brommel:2007zz}, using $N_f=2$ clover fermions at four $\beta$ values, several lattice spacings between 0.068 - 0.115 fm and a wide range of pion mass values between 440 - 1173 MeV. They report $\avgxxx^u_\pi=$0.074(9)(4) in the $\overline{\rm MS}$ scheme at 2 GeV at the physical pion mass obtained from a chiral extrapolation.

\vspace*{0.2cm}
Much more recently, there have been explorations of the $x$-dependent pion and kaon PDFs~\cite{Karpie:2018zaz,Sufian:2019bol,Izubuchi:2019lyk,Joo:2019bzr,Bali:2019dqc,Lin:2020ssv,Gao:2020ito,Sufian:2020vzb}, and some extract the first moments with indirect methods. Unlike our work, and the aforementioned calculations, these new methods calculate matrix elements of non-local operators, with the quark fields separated by a finite spatial distance connected through a straight Wilson line. Therefore, the mixing observed in the moments of PDFs from towers of $n$-derivative local operators, is not relevant here. However, these methods have other systematic uncertainties, due to the need of boosted meson states, namely either requiring the boost to be large or the product of spatial separation and boost to be large. Furthermore, a matching kernel is necessary to relate these matrix elements to the light-cone PDFs.

in Ref.~\cite{Joo:2019bzr}  two ensembles of $N_f=2+1$ clover fermions with pion mass 415 MeV and two volumes were used. They extracted the $x$-dependence of PDFs using the pseudo-ITD approach. They used two methods to extract the lowest moments: OPE without an OPE and an integration of the PDFs that gives $\avgxxx^u_\pi=0.046(19)$ in the $\overline{\rm MS}$ at 2 GeV. Ref.~\cite{Lin:2020ssv} employs the quasi-PDFs method using a mixed-action setup (clover on HISQ). The calculation was performed on three ensembles with pion mass 217, 310, 319 MeV and two lattice spacings ($a = 0.06,\, 0.12$ fm). $\avgxxx^u_\pi$ was renormalized non-perturbatively and a value of $\avgxxx^u_\pi=0.057(10)$ at a scale of 5.2 GeV is reported after chiral extrapolation to the physical point. A similar analysis for the kaon gives $\avgxxx^u_K=0.042(6)$ and $\avgxxx^u_K=0.070(6)$, but the scale is not reported. The result for the pion is presented in Table~\ref{tab:x3_pi_compare_lattice} after evolution to 2 GeV. Finally, the work of Ref.~\cite{Gao:2020ito} explored both the quasi-PDFs and pseudo-ITD approaches, and also used a mixed-action setup of clover valence fermions on $N_f=2+1$ HISQ configurations. The pion PDF was calculated using two ensembles at $m_\pi=300$ MeV and lattice spacing 0.04 and 0.06 fm. The third moment is obtained with fits to the pion PDF with a 2- and a 4-parameter Ansatz. They report the valence case $\avgxxx_\pi=0.0652(49)(36) ,\,0.0647(47)(38)$, for the 2-parameter and 4-parameter fits, respectively. The results are given in the $\overline{\rm MS}$ scheme at 3.2 GeV. We extract $\avgxxx_\pi$ at 2 GeV using the finest lattice and the 2-parameter fit, which is provided in Table~\ref{tab:x3_pi_compare_lattice}. For completeness, we summarize our calculation for the kaon in Table~\ref{tab:x3_kaon_compare_lattice}.

  \begin{table}[h]
    \begin{tabular}{|l|l|l|l|l|l|l|l|l|}
\hline\\[-2.5ex]
$\,\,$ Reference $\,\,$ & 
$\quad\,$ Method & 
\,\,\,  Renorm. \,\,   & 
$\quad$ mixing $\quad$ & 
$\,$ $m_\pi$ (MeV) & 
$\quad$ $N_f$ & 
$\,$ $\avgxxx^u_{\pi}$ (2GeV) $\,$  &  
initial scale \\[0.5ex]
\hline\\[-2.5ex]
\hspace*{0.15cm} This work &\,\,local operator\,  & \, non-perturb. &\,\, not present &$\quad$  260 &$\,\,$ 2+1+1
&\quad\,0.024(18)\,\,\,   &\quad 2 GeV\quad     \\
\hspace*{0.21cm} Ref.~\cite{Best:1997qp}  & \,\,local operator\,  & \, perturb.  &\,\,\, present  &$\,\,$   chiral extrap.  &$\,\,$ 0
&\quad\,0.051(21)   &\quad 2.4 GeV\quad       \\ 
\hspace*{0.21cm} Ref.~\cite{Detmold:2003tm} & \, local operator \, & \, perturb. \, &\,\,\, present &$\,\,$  chiral extrap. \,\,    &$\,\,$ 0  
&\quad\,0.046(16)   &\quad 2.4 GeV\quad      \\ 
\hspace*{0.21cm} Ref.~\cite{Brommel:2007zz} & \, local operator \,  & \, non-perturb. \,  &\,\,\, present &$\,\,$  chiral extrap.     &$\,\,$ 2  
&\quad\,0.074(10)\,\,\,    &\quad 2 GeV\quad       \\ 
\hspace*{0.21cm} Ref.~\cite{Joo:2019bzr}   & \, pseudo-ITD \, & \, non-perturb. \, & \,\,\,\,N/A   &$\quad$  415  &$\,\,$ 2+1 
&\quad\,0.046(19)\,\,\,    &\quad 2 GeV\quad    \\ 
\hspace*{0.21cm} Ref.~\cite{Lin:2020ssv}  & \, quasi-PDF \,  & \, non-perturb. \, & \,\,\,\,N/A  &$\,\,$   chiral extrap.  &$\,\,$ 2+1+1 &\quad\,0.073(13)  &\quad 5.2 GeV\quad    \\ 
\hspace*{0.21cm} Ref.~\cite{Gao:2020ito}   & \, pseudo-ITD \, & \, non-perturb.\,  & \,\,\,\,N/A  &$\quad$  300 &$\,\,$ 2+1  &\quad\,0.075(61) \,\,\,   &\quad 3.2 GeV\quad    \\[0.5ex]  
\hline
    \end{tabular}
        \caption{Comparison of lattice results for $\avgxxx^u_\pi$ in the $\overline{\rm MS}$ scheme at 2 GeV. The evolution from the reported scale (``initial'' scale) to 2 GeV is applied to NNLO. Statistical and systematic uncertainties have been added in quadrature where applicable.}
        \label{tab:x3_pi_compare_lattice}
  \end{table}

As can be seen in Table~\ref{tab:x3_pi_compare_lattice}, there is a range of values obtained with different methods. We find that, our results are compatible within uncertainties with the calculations of Refs.~\cite{Best:1997qp,Detmold:2003tm}, which, however,  used an operator that exhibits mixing.  Ref.~\cite{Best:1997qp} employs a perturbative renormalization prescription, which justifies the higher value compared to the other studies. The value of Refs.~\cite{Brommel:2007zz} is higher than the other calculations with local operators. The comparison with the indirect methods to extract the Mellin moments from integration or fits on the pion PDF, also shows compatibility with our value, except the one for Ref.~\cite{Lin:2020ssv}, which is at the high end. It should be noted that these calculations carry very large uncertainties ($\sim 15\% - 95\%$) and the comparison is inconclusive. From these results one can extract the range in which $\avgxxx_\pi$ is. It is worth mentioning that our value is at the lower end of the range, which is a consequence of the suppression of excited states. For example, our results for $t_s/a=14$, which is at the range used in other calculations is higher (0.031(15)). We emphasize that we obtain directly $\avgxxx$ using a local operator that avoids mixing with lower-dimensional operators, but requires all spatial components of the meson momentum boost to be nonzero. The lattice data are renormalized non-perturbatively, and the renormalization function is multiplicative.
  

  \begin{table}[h]
 \hspace*{-0.35cm}
    \begin{tabular}{|l|l|l|l|l|l|l|l|l|}
\hline\\[-2.5ex]
$\,$ Reference $\,$ & 
$\quad\,$ Method & 
$\,\,$ Renorm. $\,\,$   & 
$\quad$ mixing $\quad$ & 
$\,$ $m_\pi$ (MeV) & 
$\quad$ $N_f$ & 
$\,$ $\avgxxx^u_{K}$ (2GeV) $\,$  &
$\,$ $\avgxxx^s_{K}$ (2GeV) $\,$  &
initial scale\\[0.5ex]
\hline\\[-2.5ex]
\hspace*{0.05cm} This work & local operator  & non-perturb. &\, not present &$\quad$  260 &$\,\,$ 2+1+1
&\quad\,0.033(6)\,\,\,   &\quad\,0.073(5)\,\,\,   &\quad 2 GeV\quad    \\[0.5ex]
\hline 
    \end{tabular}
        \caption{Our lattice results for $\avgxxx^u_K$ and $\avgxxx^s_K$ in the $\overline{\rm MS}$ scheme at 2 GeV. Statistical and systematic uncertainties have been added in quadrature.}
        \label{tab:x3_kaon_compare_lattice}
  \end{table}  

\subsection{Model calculations and global fits}

There are a few model calculations and global fits on experimental data for the three-four lowest moments of pion and kaon PDFs, and we compare these here with our results. We emphasize that the comparison is only qualitative, as many of the calculations do not have quantified uncertainties. Also, our calculation is at higher-than-physical pion and kaon masses and only the connected diagram is included.

One of the first calculations is a next-to-leading-order analysis of several $\pi^\pm N$ experimental data, including Drell-Yan and prompt photon production, presented in Ref.~\cite{Sutton:1991ay}. They obtain $\avgxxx^u_\pi=0.058(4)$ at a scale of 2 GeV. Much later, an updated analysis of the moments of pion PDFs to next-to-leading-order using the Fermilab E-615 pionic Drell-Yan data was carried out and can be found in Ref.~\cite{Wijesooriya:2005ir}. A value of $\avgxxx^u_\pi=0.045(3)$ is given at 5.2 GeV. The JAM global fit analysis is performed at 1.3 GeV for the third non-trivial moment of the pion~\cite{Barry:2018ort}, and the obtained value is $\avgxxx^u_\pi=0.074$.

Ref.~\cite{Hecht:2000xa} presents a calculation of the valence quark PDF for the pion using Schwinger-Dyson equations (DSE) and obtains $\avgxxx^u_\pi=0.049$ at 2 GeV. A more recent DSE study can be found in Ref.~\cite{Chen:2016sno} with $\avgxxx^u_\pi=0.052$ at 2 GeV. Ref.~\cite{Ding:2019qlr} reports $\avgxxx^u_\pi=0.049(7)$ at 2 GeV using the Bethe-Salpeter equation (BSE). The recent calculation of Ref.~\cite{Bednar:2018mtf} applied a rainbow-ladder truncation of DSEs, and therefore all planar diagrams were summed and the non-perturbative gluon dressing of the quarks was correctly accounted for. They find $\avgxxx_\pi^u=0.109$ at 0.78 GeV. A calculation using the chiral constituent quark model is found in Ref.~\cite{Watanabe:2017pvl}, which gives $\avgxxx^u_\pi=0.048$ at a scale of 5.2 GeV. Finally, Ref.~\cite{Lan:2019rba} combined QCD evolution with light front quantization to obtain the pion PDFs, as well as the moments up to $\avgxxxx$. At 2 GeV they find $\avgxxx^u_\pi=0.057(8)$.

The kaon $\avgxxx$ was also studied in Ref.~\cite{Chen:2016sno} and the results are $\avgxxx^u_K=0.048$ and $\avgxxx^s_K=0.092$ at 2 GeV. The model calculation of Ref.~\cite{Watanabe:2017pvl} gives $\avgxxx^u_K=0.045$ and $\avgxxx^s_K=0.049$ at 5.2 GeV. The findings of Ref.~\cite{Bednar:2018mtf} are $\avgxxx_K^u=0.092$ and $\avgxxx_K^s=0.143$ at 0.78 GeV. Ref.~\cite{Lan:2019rba} report $\avgxxx^u_K=0.050(6)$ and  $\avgxxx^s_K=0.066(9)$ at 2 GeV.

The aforementioned results can be found in Table~\ref{tab:x3_pi_compare} for the pion and Table~\ref{tab:x3_kaon_compare} for the kaon, after evolution to 2 GeV. An extended list can be found in Ref.~\cite{Lan:2019rba}. We find that our data for the pion and up part of kaon are lower than most of the other calculations. However, some of the calculations do not include systematic uncertainties, which prevents a meaningful comparison. A better agreement is observed for the strange part of the kaon with our values being in the middle of the range from the other calculations. Let us remind the reader that our calculation, like all other lattice results mentioned above, focuses in the connected diagram. Also, the ensemble used has a pion mass of 260 MeV and a kaon of 530 MeV, which are higher than the physical values.

  \begin{table}[h]
    \begin{tabular}{|l|l|l|}
\hline\\[-2.5ex]
$\qquad$ Reference & $\,$ $\avgxxx^u_{\pi}$ (2GeV) $\,$  & initial scale\\[0.5ex]
\hline\\[-2.5ex]
\,\,This work           (lattice)  &\quad\,0.024(18)\,\,\,  &\quad 2 GeV\quad  \\
\,\,Ref.~\cite{Sutton:1991ay} (global fit)  &\quad\,0.058(4)\,\,\, &\quad 2 GeV\quad  \\ 
\,\,Ref.~\cite{Hecht:2000xa} (DSE) \,\,  &\quad\,0.049\,\,\, &\quad 2 GeV\quad  \\ 
\,\,Ref.~\cite{Wijesooriya:2005ir} (global fit)\,\,   &\quad\,0.058(4)\,\,\,   &\quad 5.2 GeV\quad  \\ 
\,\,Ref.~\cite{Chen:2016sno} (DSE)   &\quad\,0.052\,\,\, &\quad 2 GeV\quad  \\ 
\,\,Ref.~\cite{Watanabe:2017pvl} (CC quark model)\,\,    &\quad\,0.061 \,\,\,  &\quad 5.2 GeV\quad  \\ 
\,\,Ref.~\cite{Bednar:2018mtf} (DSE) \,\,    &\quad\,0.065 \,\,\,  &\quad 0.78 GeV\quad  \\  
\,\,Ref.~\cite{Barry:2018ort} (JAM global fit)\,\,    & \quad\,0.063 \,\,\, &\quad 1.3 GeV\quad  \\ 
\,\,Ref.~\cite{Ding:2019qlr} (DSE, BSEs)\,\,   &\quad\,0.049(7)\,\,\,  &\quad 2 GeV\quad\\ 
\,\,Ref.~\cite{Lan:2019rba} (BLFQ-NJL)\,\,   &\quad\,0.057(8)\,\,\,  &\quad 2 GeV\quad\\[0.5ex] 
\hline
    \end{tabular}
        \caption{Comparison with global fits and model calculations for $\avgxxx_\pi$ in the $\overline{\rm MS}$ scheme at 2 GeV. The evolution from the reported scale (``initial'' scale) to 2 GeV is applied to NNLO. Statistical and systematic uncertainties have been added in quadrature where applicable. }
        \label{tab:x3_pi_compare}
  \end{table}

  \begin{table}[h]
    \begin{tabular}{|l|l|l|l|}
\hline\\[-2.5ex]
$\qquad$ Reference  & $\,$ $\avgxxx^u_{K}$ (2GeV) $\,$   & $\,$ $\avgxxx^s_{k}$ (2GeV) $\,$   & initial scale \\[0.5ex]
\hline\\[-2.5ex]
\,\,This work            (lattice)  &\quad\,0.033(6)\,\,\,   &\quad\,0.073(5)\,\,\,&\quad 2 GeV\quad   \\
\,\,Ref.~\cite{Chen:2016sno} (DSE)   &\quad\,0.048\,\,\,    &\quad\,0.092\,\,\, &\quad 2 GeV\quad  \\ 
\,\,Ref.~\cite{Watanabe:2017pvl} (CC quark model) \,\,   &\quad\,0.058\,\,\,   &\quad\,0.063\,\,\, &\quad 5.2 GeV\quad  \\ 
\,\,Ref.~\cite{Bednar:2018mtf} (DSE) \,\,    &\quad\,0.55 \,\,\,  &\quad\,0.85 \,\,\,  &\quad 0.78 GeV\quad  \\ 
\,\,Ref.~\cite{Lan:2019rba} (BLFQ-NJL)\,\,   &\quad\,0.050(7) \,\,\,  &\quad\,0.066(9) \,\,\, &\quad 2 GeV\quad\\[0.5ex]
\hline
    \end{tabular}
        \caption{Comparison with global fits and model calculations for $\avgxxx^u_K$ and $\avgxxx^s_K$ in the $\overline{\rm MS}$ scheme at 2 GeV. The evolution from the reported scale (``initial'' scale) to 2 GeV is applied to NNLO. Statistical and systematic uncertainties have been added in quadrature where applicable.}
        \label{tab:x3_kaon_compare}
  \end{table}

\section{Reconstruction of PDFs}
\label{sec:PDFs}

It is generally believed that the reconstruction of PDFs from their Mellin moments is, at best, challenging on the lattice for a number of  reasons. The signal-to-noise ratio decays fast with the addition of derivatives in the operator, requiring increased computational cost so that gauge noise is controlled. The moments $\langle x^2 \rangle$ and $\langle x^3 \rangle$ can only be obtained in a kinematic framework where the hadron has momentum with nonzero spatial components. In fact, to avoid mixing with lower dimensional operators under renormalization, the initial and final states should carry momentum with at least two and three nonzero spatial  components, respectively. Such a setup comes at increased computational cost. In addition, the mixing under renormalization for the moments with $n>3$ cannot be avoided, regardless of the kinematic framework. Because of these challenges, early attempts have been inconclusive in determining whether it is feasible to reconstruct PDFs from lower moments using lattice QCD (see, e.g., Refs.~\cite{Detmold:2001dv,Holt_2010}). More recently, methods to extract higher moments have been proposed, using smeared operators~\cite{Davoudi:2012ya}, heavy-quark operator product expansion (HOPE)~\cite{Detmold:2005gg,Detmold:2021uru}, and light-quark current-current correlators~\cite{Braun:2007wv}.

While the above challenges are true, numerical simulations have advanced significantly with more computational power, better algorithms, implementation of non-perturbative renormalization, and methods to control gauge noise. In fact, in this work and in Ref.~\cite{Alexandrou:2020gxs}, we demonstrate that the Mellin moments with $n<4$ can be obtained with reliable elimination of excited states and in a setup that does not contain mixing with lower dimensional operators. Therefore, we attempt the reconstruction of the $x$-dependence of PDFs. The goal is threefold: \textbf{a.} understand the limitations of the reconstruction; \textbf{b.} study the large-$x$ behavior; and \textbf{c.} extract the moments with $n>3$ from the reconstructed PDFs. 
 
\subsection{Setup}

We use the standard functional form
\begin{equation}
\label{eq:PDF}
q^f_M(x) = N x^\alpha (1-x)^\beta (1+\rho \sqrt{x} + \gamma x)   \,, 
\end{equation} 
to obtain the $x$ dependence of the pion and kaon PDFs, $q^f_M(x)$, where $(M,f)=(\pi,\,u),\,(K,\,u),\,(K,\,s)$. $N$ is a normalization defined by charge conservation
\begin{equation}
\langle 1 \rangle_M = \int_0^1 q_M(x) = 1\,,
\end{equation} 
leading to
\begin{equation}
\label{eq:Norm}
N= \frac{1}{B(\alpha+1,\beta+1) + \gamma B(2+\alpha,\beta+1) }\,,
\end{equation} 
where $B$ is the Euler beta-function. The fit parameters in Eqs.~(\ref{eq:PDF}), (\ref{eq:Norm}) are $\alpha$, $\beta$, $\gamma$ and $\rho$. Their values depend on $M$ and $f$, but we omit the subscript and superscript in the following equations for simplicity in the notation. The parameter $\rho$ is generally assumed to be negligible~\cite{Chen:2016sno}, and therefore, we omit the term $\rho \sqrt{x}$. By integrating Eq.~(\ref{eq:PDF}), we can extract the $n^{\rm th}$-moment as a function of the fit parameters, that is
\begin{equation}
\label{eq:moments}
\displaystyle
\langle x^n \rangle = \frac{ \Big{(}\prod_{i=1}^n (i+\alpha) \Big{)} \, \Big{(}n+2+\alpha+\beta+(i+1+\alpha) \gamma\Big{)}}
{\Big{(}\prod_{i=1}^n (i+2+\alpha+\beta) \Big{)}\, \Big{(}2+\alpha+\beta+(1+\alpha)\gamma\Big{)}} \,,\quad n>0\,.
\end{equation}
Our results for $\langle x^n\rangle$, $n=1,2,3$ are used as input for Eq.~(\ref{eq:moments}) to extract the fit parameters. The results of $\avgx$ are given in Eqs.~(38) - (40) of Ref.~\cite{Alexandrou:2020gxs}. For $\avgxx$ we use the values given in Eqs.~(\ref{eq:x2_p_u}) - (\ref{eq:x2_k_s}), which have been obtained at higher statistics than the values reported in Ref.~\cite{Alexandrou:2020gxs}. $\avgxxx$ is given in Eqs.~(\ref{eq:x3_p_u}) - (\ref{eq:x3_k_s}). To compare with results from global fits and models, we evolve our results for the moments to a scale of 5.2 GeV using NNLO expressions.

\subsection{Lattice data}

We apply a 2- and a 3-parameter fit to examine the effects on the PDF reconstruction. In the case of the 2-parameter fit we set $\gamma=0$, as  used in many such fits. The values extracted for the parameters from the fits are given in Table~\ref{table:fit_param_pion_u} for the pion and kaon. We find that the fits for the pion are less stable than the ones for the kaon, due to the enhanced gauge noise in the former. For the case of the pion and the strange contribution of the kaon, we find that the parameter $\gamma$ has large uncertainties. This is due to the fact that the number of moments are not enough to perform a 3-parameter fit, and the parameters have a competitive role in the fit. For instance, we find that in the pion and strange kaon, $\gamma=0$ within errors. 

\begin{table}[h!]
    \begin{tabular}{ccccc}
      \hline\hline\\[-2.5ex]
 fit type & $\qquad\alpha^u_{\pi}\qquad$ & $\qquad\beta^u_{\pi}\qquad$   & $\qquad\gamma^u_{\pi}\qquad$ & $\qquad\chi^2/\rm{d.o.f.}\qquad$\\[0.5ex]
\hline\\[-2.5ex]
2-parameter  &$\,\,\,$-0.04(20)   &$\,\,\,$2.23(65)     &$\,\,\,$0            & $\,\,\,$1.48  \\
3-parameter  &$\,\,\,$-0.54(22)   &$\,\,\,$2.76(64)     &$\,\,\,$22.17(17.87) & $\,\,\,$2.77 \\ [0.5ex]
      \hline\hline \\[-2.5ex]
 fit type & $\qquad\alpha^u_{K}\qquad$ & $\qquad\beta^u_{K}\qquad$   & $\qquad\gamma^u_{K}\qquad$ & $\qquad\chi^2/\rm{d.o.f.}\qquad$\\[0.5ex]
\hline\\[-2.5ex]
2-parameter  &$\,\,\,$-0.05(7)   &$\,\,\,$ 2.42(24)     &$\,\,\,$0            & $\,\,\,$2.21     \\
3-parameter  &$\,\,\,$ -0.56(72)   &$\,\,\,$3.01(23)     &$\,\,\,$25.11(5.23) & $\,\,\,$3.78     \\ [0.5ex]
      \hline\hline\\[-2.5ex]
 fit type& $\qquad\alpha^s_{K}\qquad$ & $\qquad\beta^s_{K}\qquad$   & $\qquad\gamma^s_{K}\qquad$ & $\qquad\chi^2/\rm{d.o.f.}\qquad$\\[0.5ex]
\hline\\[-2.5ex]
2-parameter  &$\,\,\,$0.21(8)   &$\,\,\,$2.13(20)     &$\,\,\,$0                & $\,\,\,$0.0036     \\
3-parameter  &$\,\,\,$0.18(95)   &$\,\,\,$2.051(3.46)     &$\,\,\,$0.347(16.10) & $\,\,\,$0.0018    \\ [0.5ex]
      \hline\hline
    \end{tabular}
        \caption{The values for the fit parameters, $\alpha,\,\beta$ and $\gamma$ for $q_\pi^u$,  $q_K^u$ and  $q_K^s$ at 5.2 GeV. The error in the parenthesis is statistical.}
        \label{table:fit_param_pion_u}
  \end{table}

\begin{figure}[!h]
    \centering
    \includegraphics[scale=0.45]{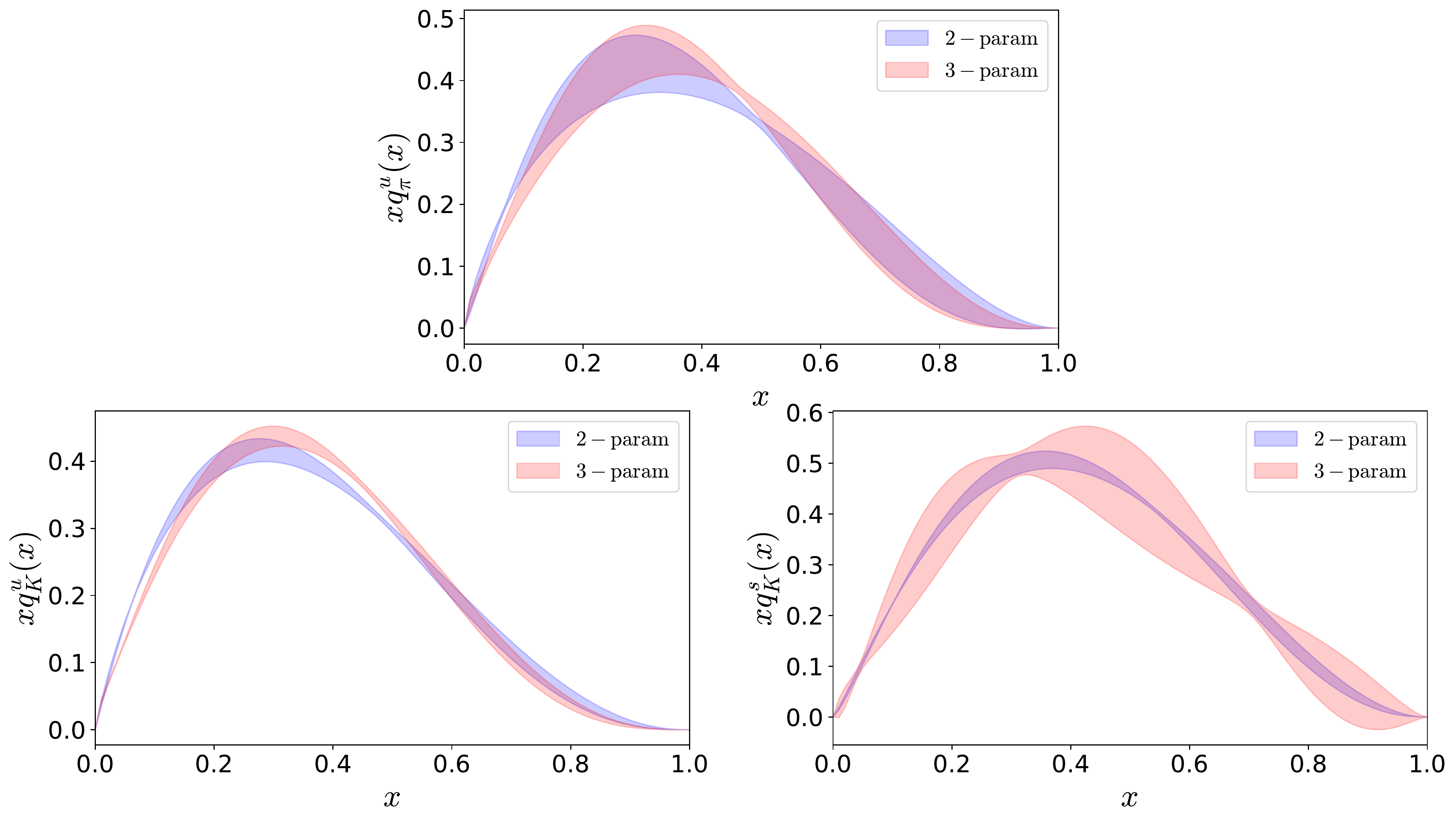}
    \vspace*{-0.2cm}
    \caption{Top: The $x$ dependence of $x q_\pi^u(x)$ at 27 GeV$^2$ for the 2-parameter (blue band) and 3-parameter (pink band) fits. Bottom: Same as top panel for $x q_K^u(x)$ (left) and  $x q_K^s(x)$ (right). Results are given in the $\overline{\rm MS}$ scheme at 27 GeV$^2$.}
    \label{fig:PDF_param}
\end{figure}

Using the parameters of Table~\ref{table:fit_param_pion_u} obtained from both fits we reconstruct the $x$-dependent PDFs as shown in Fig.~\ref{fig:PDF_param}. We find that the shape of the PDFs has a mild dependence on the choice of fit, mostly in $q^{u}_K$. The uncertainties for $q^{s}_K$ are increased for the 3-parameter fit, while they are very similar for the other two PDFs. For the remaining presentation we focus on the 2-parameter fits.

\begin{figure}[!h]
    \centering
    \includegraphics[scale=0.45]{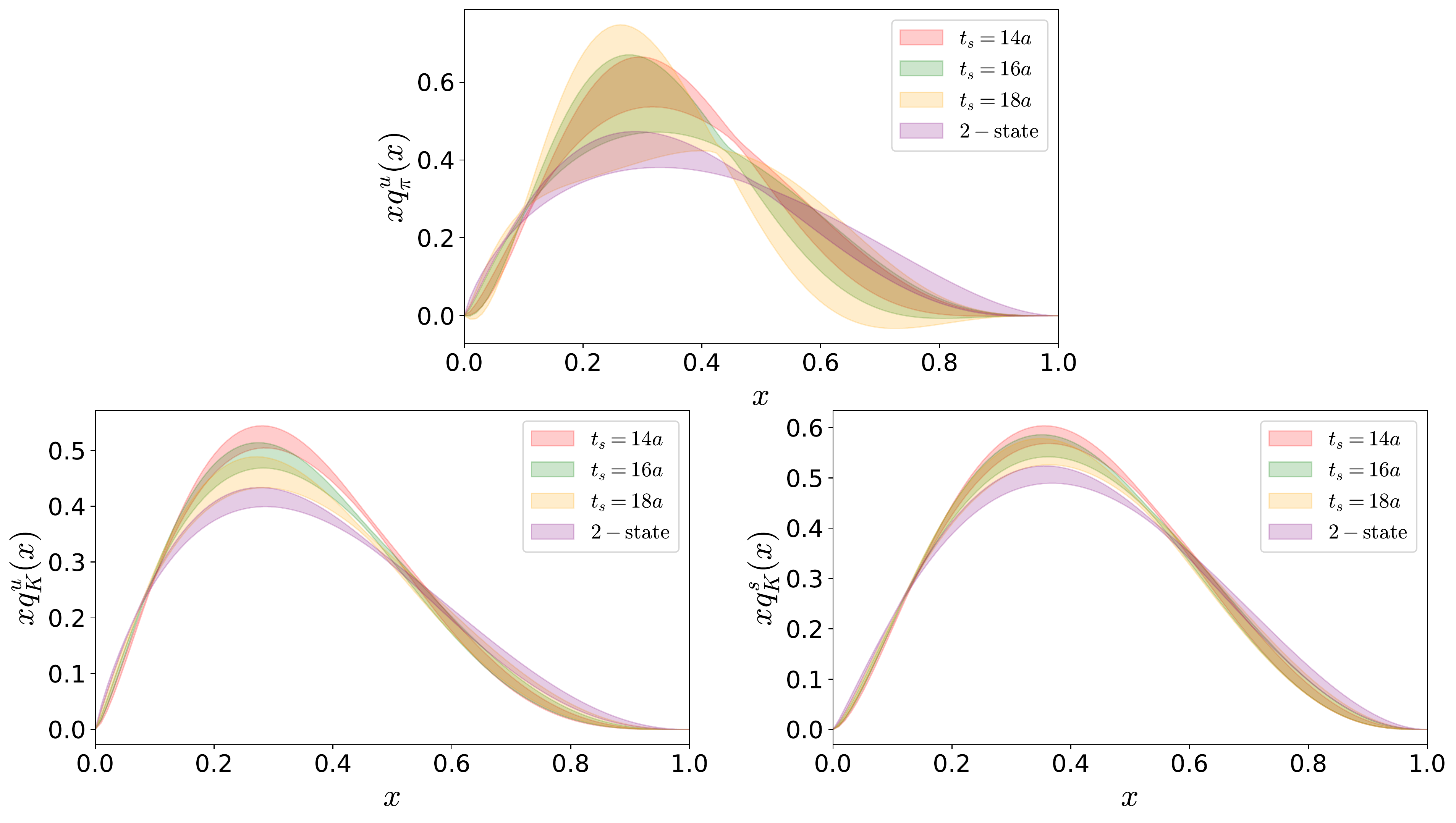}
    \vspace*{-0.2cm}
    \caption{Dependence of $x q_\pi^u(x)$ (top), $x q_K^u(x)$ (lower left panel) and $x q_K^s(x)$ (lower right panel) on the source-sink time separation. The pink, green, yellow and purple bands correspond to $t_s/a=$14, 16, 18 and the 2-state fit, respectively. The results are given in the $\overline{\rm MS}$ scheme at a scale of 27 GeV$^2$.}
    \label{fig:xPDF_tsink}
\end{figure}

Next, we want to study the effects of excited-states contamination on $q_M^f(x)$ that may be non-trivial because the dependence of the fit parameters in Eq.~(\ref{eq:PDF}) on the moments is non-linear. We  apply the 2-parameter fit on our results at $t_s/a=14 - 18$, as well as the two-state fit and reconstruct the PDFs, as shown in Fig.~\ref{fig:xPDF_tsink}. We observe a nice convergence for all the cases as $t_s$ increases. Similarly to the  behavior of the moments, we find that the excited-states contamination at $t_s \lesssim 1.4$ fm leads to a PDF that is higher than the two-state fit and the plateau fit at $t_s \geqslant 1.5$ fm. Eventually, the two-state fit values converge with a peak around $x\sim 0.3 - 0.4$, that is $x q_\pi^u(0.3)\sim 0.4$, $x q_K^u(0.3)\sim 0.4$ and $x q_K^s(0.4)\sim 0.5$. We use as final PDFs those extracted from a 2-parameter fit on the two-state fit results for the moments (purple band in Fig.~\ref{fig:xPDF_tsink}). 

\begin{figure}[!h]
    \centering
    \includegraphics[scale=0.45]{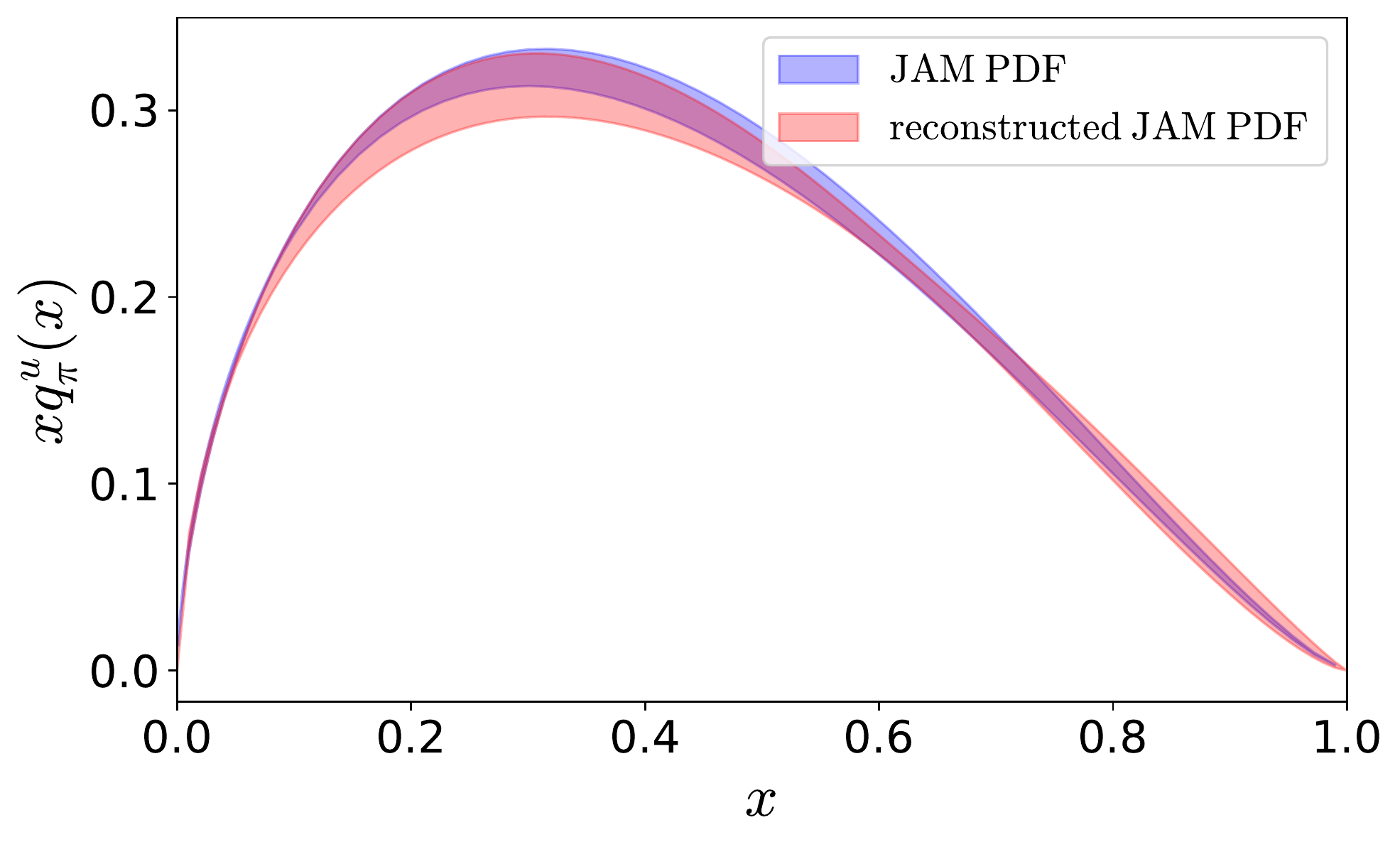}
    \vspace*{-0.5cm}
    \caption{Pion PDF using the JAM data (blue band) and the reconstructed PDF using its moments with $n \leq 3$ (pink band). The reported scale is 27 GeV$^2$.}
    \label{fig:PDF_JAM}
\end{figure}

One concern is whether the use of the $n \leq 3$ moments can successfully reconstruct the PDF. To address this question we use the JAM data for the pion PDF~\cite{Barry:2018ort}, as well as its moments. We follow the same procedure as for our lattice data, that is, fit the parameters $\alpha$, $\beta$, $\gamma$ using as input only the JAM moments with $n \leq 3$, and then produce the reconstructed PDF via Eq.~(\ref{eq:PDF}) with $\rho=0$. The reconstructed PDF is then compared to the original JAM PDF and is shown in Fig.~\ref{fig:PDF_JAM}. We use bootstrap sampling to obtain the uncertainties and width of the band. As can be seen, the two are in very good agreement within uncertainties for almost all $x$ region. The reconstruction of the PDF from its moments with only $n \leq 3$ leads to much larger uncertainties due to the truncation of the information that is used to extract the PDF. Furthermore, we use the reconstructed PDF to estimate the $n=4$ moment via Eq.~(\ref{eq:moments}), and we find $\avgxxxx_\pi^u=0.026(2)$. This is in excellent agreement with the moment as extracted from the JAM framework, $\avgxxxx_\pi^u=0.027(2)$, which is more accurate, as expected. 

\begin{figure}[!h]
    \centering
    \includegraphics[scale=0.45]{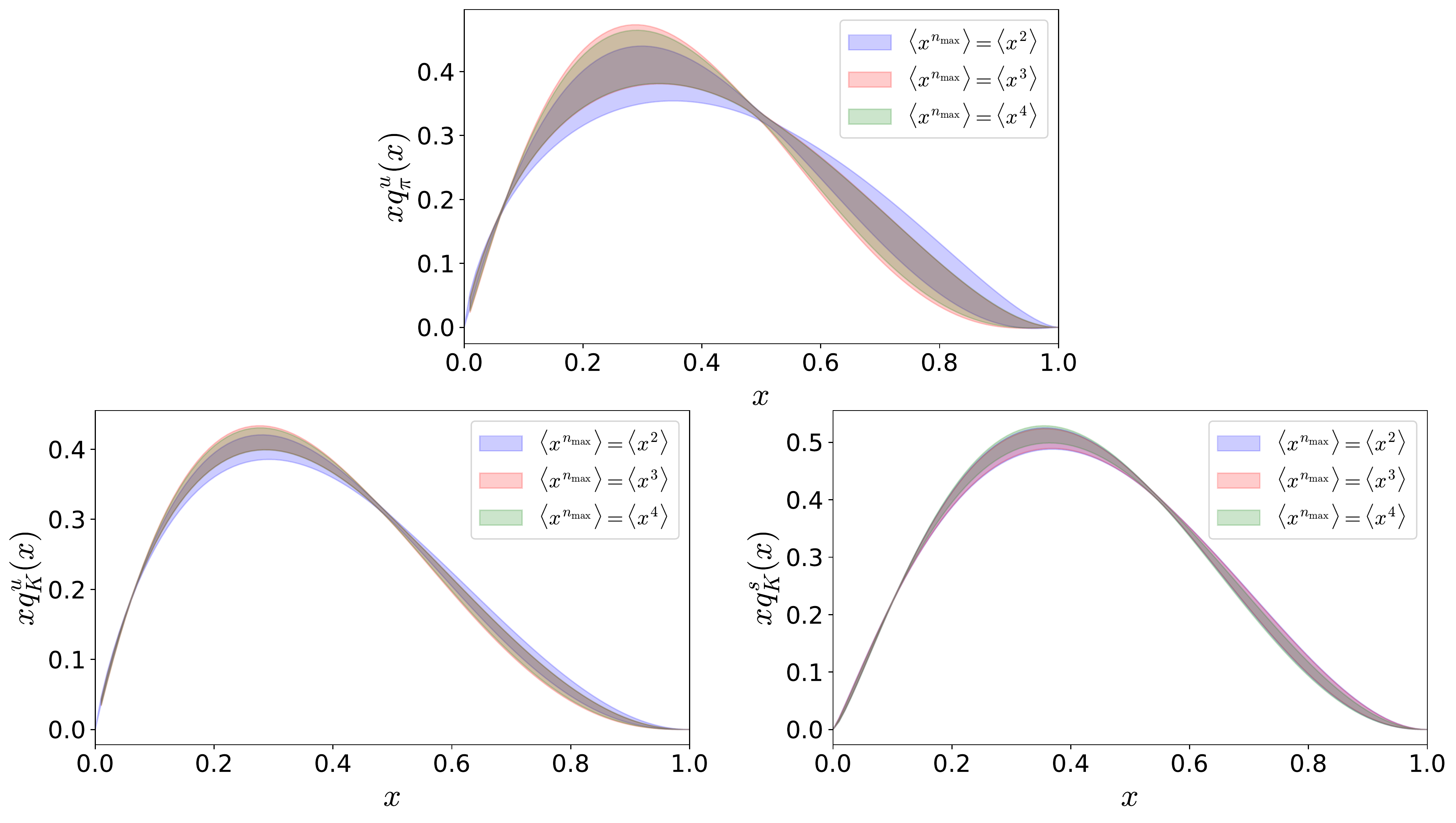}
    \caption{Top: The $x$ dependence of $x q_\pi^u(x)$ using the 2-parameter with $\langle x^n \rangle_{\rm max}=\avgxx,\,\avgxxx,\,\avgxxxx$, shown with blue, pink and green bands. For the green band we use a constraint of $\avgxxxx_\pi^u$~\cite{Barry:2018ort}. Bottom: Same as top panel for $x q_K^u(x)$ (left panel) and $x q_K^s(x)$ (right panel) using the BLFQ-NJL~\cite{Lan:2019rba} $\avgxxxx_K$ values as constraints for the pink band.}
    \label{fig:PDF_x4_constraint}
\end{figure}

These results suggest that the reconstruction of the PDFs using the $n \leq 3$ moments is indeed feasible. To further test the sensitivity of our fits, we change the number of inputs used. In particular, we perform fits including moments up to $\langle x^{n_\textrm{max}} \rangle$, with $n_\textrm{max}=$2, 3, or 4. For $n_\textrm{max}=$2 and 3 we only use lattice data, while for $n_\textrm{max}=$4 we add another constraint by using the value of $\avgxxxx$ from global fits and models for the pion and kaon, respectively. In particular, we use $\avgxxxx_\pi^u=0.027(2)$ from the JAM analysis~\cite{Barry:2018ort}, and $\avgxxxx_K^s=0.029^{+0.005}_{-0.004}$, $\avgxxxx_K^u=0.021^{+0.003}_{-0.003}$ from BLFQ-NJL~\cite{Lan:2019rba}. It should be emphasized that, combining lattice data with results from model calculations is a useful exercise for understanding the effect on the fits. However, it is not a preferred direction due to the various sources of uncertainties, and only lattice data enter our final results. We remind the reader that our calculation is performed at a pion mass of 260 MeV and a kaon of 530 MeV, which are larger than their physical values. While for the kaon this is only 7$\%$ larger, for the pion this corresponds to a factor of about two. However, it was shown that the pion moment $\avgx$ has insignificant pion mass dependence~\cite{Abdel-Rehim:2015owa}, which implies the same for the higher moments. Therefore, one can neglect the pion mass dependence when combining our results with the JAM moment $\avgxxxx$. In Fig.~\ref{fig:PDF_x4_constraint} we compare the resulting PDF ($\langle x^{n_\textrm{max}} \rangle=\avgxxxx$) to the one that uses only lattice data ($\langle x^{n_\textrm{max}} \rangle=\avgxx,\,\avgxxx$). The results show that the addition of $n=3$ improves the constraint of PDFs. Interestingly, the addition of $n=4$ does not affect the shape of the PDFs. Therefore, the effect of higher moments is within the shown uncertainties. To summarize, our final estimates are obtained from a 2-parameter fit applied on the two-state fit results using our lattice results up to $\avgxxx$.

The large-$x$ behavior of the pion and kaon PDFs has been of great interest, due to different findings between existing data and model calculations. For the pion, the analysis of the pion Drell-Yan data from the Fermilab E615 experiment~\cite{Conway:1989fs} suggests a $(1-x)^1$ fall ($\beta=1$), while the data of Ref.~\cite{Aicher:2010cb} indicate a $(1-x)^2$ dependence ($\beta=2$). Also, DSE results~\cite{Chen:2016sno} find a coefficient $\beta$ closer to 2, in support of the arguments that the distribution at large-$x$ is dominated by the term $(1-x)^{2+\overline{\gamma}}$, where $\overline{\gamma}\geq 0$ is an anomalous dimension. One can argue either direction: Ref.~\cite{Conway:1989fs} uses evolution equations to only leading order in perturbative QCD, which is not sufficient for convergence. On the other hand, the model calculations have limitations on how well they describe QCD. To date, the tension persists. Therefore, it is desirable to address this issue from lattice QCD. Recently, the large-$x$ dependence was discussed using new methods to access the $x$-dependence of PDFs (see, e.g., Refs.~\cite{Chen:2018fwa,Izubuchi:2019lyk,Joo:2019bzr,Sufian:2019bol,Lin:2020ssv,Sufian:2020vzb}). In this work, we address the large-$x$ behavior using the reconstructed PDFs from their Mellin moments. As can be seen in Table~\ref{table:fit_param_pion_u} and in Fig.~\ref{fig:PDF_param}, our data show a preference in the functional form $\sim(1-x)^2$ for both the pion and kaon. 

\begin{figure}[!h]
    \centering
    \includegraphics[scale=0.5]{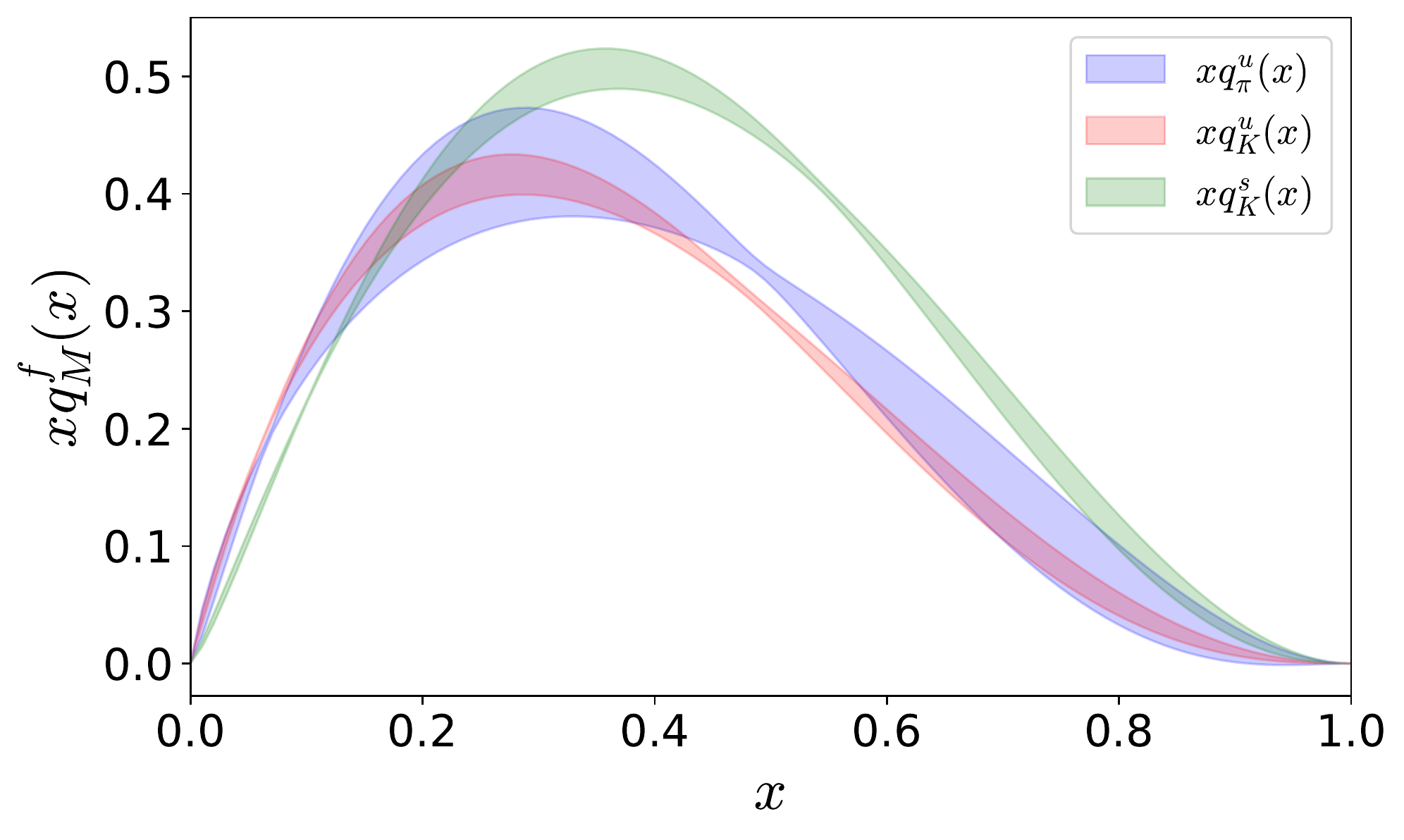}
    \caption{Comparison of $x q_\pi^u(x)$ (blue band), $x q_K^u(x)$ (pink band) and $x q_K^s(x)$ (green band) at 27 GeV$^2$. The reconstruction uses our lattice data up to $\avgxxx$ obtained with the 2-state fits analysis and a 2-parameter fit.}
    \label{fig:xPDF}
\end{figure}

The pion and kaon PDFs can be compared in order to address SU(3) flavor symmetry, which is broken due to the larger mass of the strange quark as compared to the up and down quark. This leads to a mass difference between the pion and kaon, a manifestation of the SU(3) flavor symmetry breaking effect in the Nambu-Goldstone bosons. To this end, we compare the pion and kaon PDFs in Fig.~\ref{fig:xPDF}. The distributions $x q_\pi^u(x)$ and $x q_K^u(x)$ are in full agreement for all regions of $x$, besides a minor tension around $x=0.5$. Based on this behavior, one can argue that the up quark plays an equal role in the pion and kaon PDF, and has most of its support in the small- to intermediate-$x$ regions. The strange quark in the kaon shows a tension with the up quark between $x=0.3$ and $x=0.8$, with the strange quark having more support in the large-$x$ region, as one would expect from quark mass effects. 
We find that the peak of the distributions are $x q_\pi^u(x=0.3)=0.43(5)$, $x q_\pi^u(x=0.28)=0.42(2)$, and $x q_\pi^u(x=0.36)=0.51(2)$.

Using the reconstructed pion and kaon PDFs we apply the appropriate integrals to extract their moments. One of the advantages of extracting the moments beyond $n=3$ from the $x$-dependent PDF is that it avoids the operator mixing problem, which one has to deal with in the calculation of matrix elements using $n^{\rm th}$-derivative operators. The moments up to $\langle x^6 \rangle$ in the $\overline{\rm MS}$ scheme at 27 GeV$^2$ are given in Table~\ref{table:fit_moments}. Namely, we use Eq.~(\ref{eq:moments}) and the parameters obtained for the 2-parameter fit tabulated in Table~\ref{table:fit_param_pion_u}. We propagate both the statistical error and the systematic error of residual excited states contamination as given in the first and second parenthesis, respectively, on the mean values in Table~\ref{table:fit_moments}. Note that the statistical uncertainties are well controlled for $\langle x^n \rangle$ with $n>3$, something that would not be feasible had these moments been calculated directly as matrix elements of the nucleon with the same statistics. Our value for $\avgxxxx$ is in agreement JAM value $\avgxxxx_\pi^u=0.027(2)$.

 \begin{table}[h!]
    \begin{tabular}{ccccccc}
      \hline\hline\\[-2.5ex]
$q_M^f$ & $\qquad\avgx\qquad$ & $\qquad\avgxx\qquad$   & $\qquad\avgxxx\qquad$ & $\qquad\avgxxxx\qquad$ & $\qquad\avgxxxxx\qquad$   & $\qquad\langle x^6 \rangle\qquad$  \\[0.5ex]
\hline\\[-2.5ex]
$q_\pi^u$  &$\,\,\,$0.230(3)(7)   &$\,\,\,$0.087(5)(8)     &$\,\,\,$0.041(5)(9)   &$\,\,\,$0.023(5)(6)   &$\,\,\,$0.014(4)(5)     &$\,\,\,$0.009(3)(3)  \\
$q_K^u$  &$\,\,\,$0.217(2)(5)   &$\,\,\,$0.079(2)(1)     &$\,\,\,$0.036(2)(2)   &$\,\,\,$0.019(1)(2)   &$\,\,\,$0.011(1)(2)     &$\,\,\,$0.007(1)(1) \\
$q_K^s$  &$\,\,\,$0.279(1)(5)   &$\,\,\,$0.115(2)(6)     &$\,\,\,$0.058(2)(2)   &$\,\,\,$0.033(2)(2)   &$\,\,\,$0.021(1)(2)     &$\,\,\,$0.014(1)(2) \\[0.5ex]
      \hline\hline
    \end{tabular}
        \caption{Values for the first six moments for the pion and kaon in the $\overline{\rm MS}$ at 27 GeV$^2$. The number shown in the first parenthesis is the statistical error, while the number in the second parenthesis is the systematic error indicating residual  excited-states contamination.}
        \label{table:fit_moments}
  \end{table}

\subsection{Comparison with other studies}

\begin{figure}[!h]
    \centering
    \includegraphics[scale=0.45]{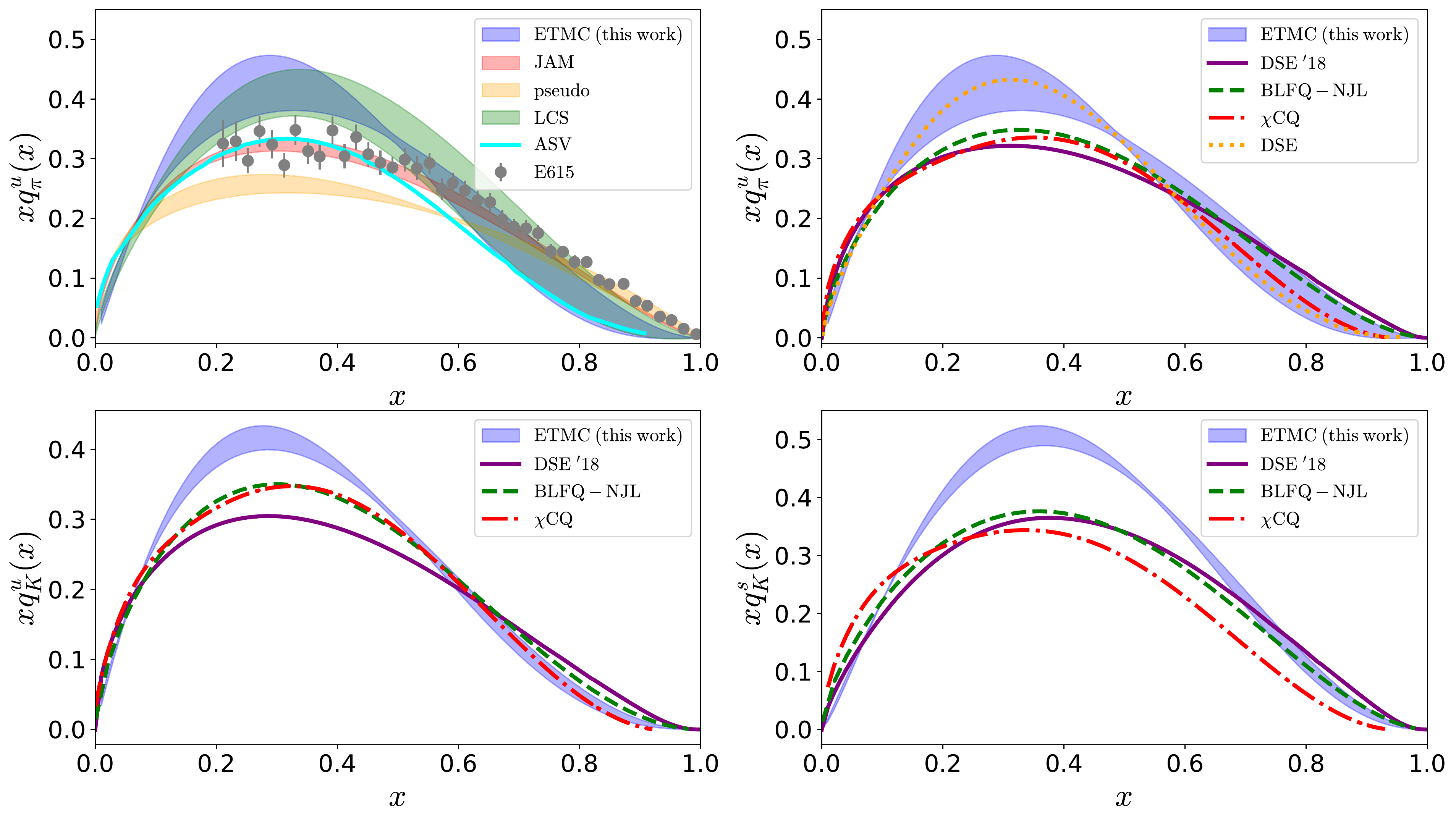}
\caption{Top left panel: Comparison of the pion $x q_\pi^u(x)$ with other lattice calculations, experimental data and global analysis, all in the $\overline{\rm MS}$ at 27 GeV$^2$. Our results (blue band) use data up to $\avgxxx$ obtained with the two-state fits analysis and a 2-parameter fit. The E615 data~\cite{Conway:1989fs} are shown with gray points, the rescaled ASV curve~\cite{Aicher:2010cb} with cyan color. The JAM global fit is shown with pink band, and the lattice results from pseudo-ITD~\cite{Joo:2019bzr} and current current correlators (LCS)~\cite{Sufian:2020vzb} are shown with orange and green band, respectively. Top right panel: Comparison of our results for $x q_\pi^u(x)$ with DSE~\cite{Chen:2016sno} and the updated DSE'18~\cite{Bednar:2018mtf}, BLFQ-NJL~\cite{Lan:2019rba} and $\chi CQ$~\cite{Watanabe:2017pvl}. Bottom panel: Same as top panel for the kaon $x q_K^u(x)$ (left) and $x q_K^s(x)$ (right).}
    \label{fig:PDF_all_fits}
\end{figure}

As mentioned in the previous section, there are a few calculations of the pion and kaon PDFs, which we compare to our results in Fig.~\ref{fig:PDF_all_fits}. In the left upper panel, we compare with the lattice results extracted from the pseudo-ITD approach~\cite{Joo:2019bzr} and the current-current correlators (LCS) method~\cite{Sufian:2019bol,Sufian:2020vzb}. We find  best  agreement with the pion PDF obtained by three parameters fit using the LCS method, while the pseudo-ITD data have a lower peak. We include in the plot  the E615~\cite{Conway:1989fs} and ASV data~\cite{Aicher:2010cb}. The latter include soft-gluon resummation and use the next-to-leading order formalism. As can be seen, while the E615 and ASV data are in agreement in the intermediate $x$-rang the ASV data fall off faster. The global fits of the JAM Collaboration~\cite{Barry:2018ort,cao20213dimensional} is also shown. The JAM fit describes well the E615 data. In the upper right panel we compare with Dyson-Schwinger (DSE)~\cite{Chen:2016sno} as well as  the updated DSE '18~\cite{Bednar:2018mtf}, the chiral constituent quark model ($\chi$CQ)~\cite{Watanabe:2017pvl}, and the BLFQ Collaboration results from in the light front quantization and QCD evolution (NJL)~\cite{Lan:2019rba}. We note that the pion PDF can also be extracted from the determination of the nucleon PDF from light-front holographic QCD~\cite{deTeramond:2018ecg}. For the kaon, there exist limited calculations. In the lower panel of Fig.~\ref{fig:PDF_all_fits} we compare with the $\chi$CQ results~\cite{Watanabe:2017pvl}, the BLFQ-NJL data~\cite{Lan:2019rba}, and DSE '18~\cite{Bednar:2018mtf} for the up quark (left) and strange quark (right). All results are in agreement in the small-$x$ region ($x < 0.1$) for the pion and $x q_K^u(x)$. The $\chi$CQ have a different slope in the small-$x$ region. For $x>0.6$ our results agree with all other results except the original E615 data. The intermediate region reveals disagreement between the various methods; our pion results agree with DSE~\cite{Chen:2016sno} for all regions of $x$, but overestimate the peak as compared to DSE'18. For the kaon we find that there is qualitative agreement in the small- and large-$x$ regions, while there is a tension in the intermediate-$x$ region. in all cases, our dta are larger in the intermediate $x$ region. However, we need to stress that there are no experimental data, and therefore, the comparison is qualitative, as all calculations carry non-quantified systematic uncertainties. 

\section{Summary\label{sec:VII}} 

In this paper we present a calculation of the Mellin moment $\avgxxx$ for the pion and kaon. We use one $N_f=2+1+1$ ensemble corresponding to a pion mass of 260 MeV and a kaon mass of 530 MeV. A momentum-boosted kinematical frame is required to access $\avgxxx$ from Eq.~(\ref{decomp2}). In fact, the momentum boost should have all spatial components nonzero to avoid mixing under renormalization. Here we use the minimum momentum, that is $|\vec{p}_i|=\frac{\sqrt{12}\pi}{L}$ (0.72 GeV). We find that this momentum is small enough and momentum smearing~\cite{Bali:2016lva} does not have an advantage. Excited-states contamination are studied using four values of $t_s$ ranging from 1.12~fm to 1.67~fm. We perform a single-state and a two-state fit to ensure ground state dominance. Our analysis shows that $t_s > 1.5$ fm is sufficient to suppress excited states. 

Within this work, we also calculate non-perturbatively the renormalization function of the three-derivative operator with all Dirac indices unequal, a choice that avoids mixing. We use the RI scheme, and then convert to the $\overline{\rm MS}$ scheme and apply evolution to 2 GeV using the RGI intermediate scheme. The results for $\avgxxx$ are given in Eqs.~(\ref{eq:x3_p_u}) - (\ref{eq:x3_k_s}) in the $\overline{\rm MS}$ at 2 GeV. We also study ratios of moments including the moments for the kaon over pion, which are connected to SU(3) flavor symmetry breaking. We find that for the low moments this breaking is 5-10$\%$, and  increases up to about 30-50$\%$ for $\avgxxx$. Such an effect indicates that the strange quark in the kaon has its support at higher $x$ values than the up quark in the pion and kaon.

One of the interesting aspects of this work is the reconstruction of the pion and kaon PDFs from the Mellin moments up to $\avgxxx$, as described in Section~\ref{sec:renorm}. We apply 2-parameter and 3-parameter fits using the standard functional form for PDFs. We examine excited-states contamination and find that a two-state fit is necessary to suppress excited state effects which are not negligible for source-sink time separation below 1.5 fm. Furthermore, we reconstruct the PDFs varying the highest moment used as input to be $\avgxx$, $\avgxxx$, or $\avgxxxx$. We find that, including $\avgxxxx$ does not improve the PDF reconstruction. 

In conclusion, we find that a 2-parameter fit of the form $x^\alpha (1-x)^\beta$ using the moments up to $\avgxxx$ is sufficient to reconstruct the PDF. We utilize our results on the PDFs in more than one ways. First, there has been a great interest on the value of $\beta$ which captures the large-$x$ behavior. Our lattice data exhibit a behavior of $\sim(1-x)^2$. Second, having the functional form of the PDFs, we can obtain the higher moments, and provide results up to $\avgxxxxxx$. Last, but very importantly, we examine the SU(3) flavor symmetry breaking in the PDFs, and find that the conclusions are consistent to those drawn when comparing ratios of the Mellin moments as described above. In particular, we see that the up quark has approximately the same contribution in the pion and kaon and that the strange quark has a more prominent contribution in the intermediate- to large-$x$ region.

 \begin{acknowledgements}
 
We would like to thank all members of ETMC for a very constructive and enjoyable collaboration. We are grateful to Patrick Barry (JAM Collaboration), Jiangshan Lan and James Vary (BLFQ Collaboration), Raza Suffian (JLAB/W$\&$M), and Akira Watanabe for providing their data for the comparisons presented.
M.C. and C.L. acknowledge financial support by the U.S. Department of Energy Early Career Award under Grant No.\ DE-SC0020405. 
S.B. is supported by the H2020 project PRACE 6-IP (grant agreement No 82376) and the EuroCC project (grant agreement No. 951740).
K.H. is financially supported by the Cyprus Research Promotion foundation under contract number POST-DOC/0718/0100 and the EuroCC project.
This work was in part supported by the U.S. Department of Energy, Office of Science, Office of Nuclear Physics, contract no.~DE-AC02-06CH11357.
This work used computational resources from Extreme Science and Engineering Discovery Environment (XSEDE), which is supported by National Science Foundation grant number TG-PHY170022. 
It also includes calculations carried out on the HPC resources of Temple University, supported in part by the National Science Foundation through major research instrumentation grant number 1625061 and by the US Army Research Laboratory under contract number W911NF-16-2-0189. 
\end{acknowledgements}

\bibliography{references.bib}

\end{document}